\documentclass{svjour3}

\usepackage{setspace}
\usepackage{hyperref}
\usepackage{tikz}
\usetikzlibrary{arrows}
\usepackage{url}
\usepackage[T1]{fontenc}
\usepackage[utf8]{inputenc}
\usepackage{enumerate, setspace, graphicx, amsmath,amssymb,amsfonts}
\usepackage{amsmath, setspace, amsfonts, amssymb, graphics,multirow}
\usepackage{dsfont}
\usepackage{color}
\usepackage[tt]{titlepic}
\usepackage{stmaryrd}
\usepackage{array} 
\usepackage{fancyhdr} 
\usepackage[export]{adjustbox} 
\usepackage{float}
\usepackage{bm}
\usepackage{txfonts}
\usepackage{xcolor}
\usepackage{ulem}
\usepackage{cancel}
\usepackage{enumitem}
\usepackage{tabularx}

\tikzstyle{int}=[draw, line width = 0.5mm, minimum size=5em]

\newcommand\clearrow{\global\let\rowmac\relax}
\clearrow

\usepackage[left=3cm,right=3cm,top=2cm,bottom=2cm]{geometry}

\begin{document}

\title{Inference in Gaussian state-space models with mixed effects for multiple epidemic dynamics}

\date{}


\author{Romain Narci \and Maud Delattre \and Catherine Larédo \and Elisabeta Vergu}

%
\institute{MaIAGE, INRAE, Université Paris-Saclay, 78350 Jouy-en-Josas, France \\ \\ R. Narci \\ \email{romain.narci@inrae.fr}  \\ \\
M. Delattre \\ \email{maud.delattre@inrae.fr}  \\ \\
C. Larédo \\ \email{catherine.laredo@inrae.fr}  \\ \\
E. Vergu \\ \email{elisabeta.vergu@inrae.fr}}

\maketitle

\begin{abstract}
The estimation from available data of parameters governing epidemics is a major challenge. In addition to usual issues (data often incomplete and noisy), epidemics of the same nature may be observed in several places or over different periods. The resulting possible inter-epidemic variability is rarely explicitly considered. Here, we propose to tackle multiple epidemics through a unique model incorporating a stochastic representation for each epidemic and to jointly estimate its parameters from noisy and partial observations. By building on a previous work, a Gaussian state-space model is extended to a model with mixed effects on the parameters describing simultaneously several epidemics and their observation process. An appropriate inference method is developed, by coupling the SAEM algorithm with Kalman-type filtering. Its performances are investigated on SIR simulated data. Our method outperforms an inference method separately processing each dataset. An application to SEIR influenza outbreaks in France over several years using incidence data is also carried out, by proposing a new version of the filtering algorithm. Parameter estimations highlight a non-negligible variability between influenza seasons, both in transmission and case reporting. The main contribution of our study is to rigorously and explicitly account for the inter-epidemic variability between multiple outbreaks, both from the viewpoint of modeling and inference.
\end{abstract}


\section{Introduction}
\label{intro}

Estimation from available data of model parameters describing epidemic dynamics is a major challenge in epidemiology, especially contributing to better understand the mechanisms underlying these dynamics and to provide reliable predictions. Epidemics can be recurrent over time and/or occur simultaneously in different regions. For example, influenza outbreaks in France are seasonal and can unfold in several distinct regions with different intensities at the same time. This translates into a non-negligible variability between epidemic phenomena. In practice, this inter-epidemic variability is often omitted, by not explicitly considering specific components for each entity (population, period). Instead, each data series is analysed separately and this variability is estimated empirically. Integrating in a unique model these sources of variability allows to study simultaneously the observed data sets corresponding to each spatial (e.g. region) or temporal entity (e.g. season). This approach should improve the statistical power and accuracy of the estimation of epidemic parameters as well as refine knowledge about underlying inter-epidemic variability.  \\

\noindent An appropriate framework is represented by the mixed-effects models, which allow to describe the variability between subjects belonging to a same population from repeated data (see e.g. \cite{Pinheiro2000}, \cite{Lavielle2014}). These models are largely used in pharmacokinetics with intra-population dynamics usually modeled by ordinary differential equations (ODE) and, in order to describe the differences between individuals, random effects on the parameters ruling these dynamics (see e.g. \cite{Collin2020}). This framework was later extended to models defined by stochastic differential equations incorporating mixed effects in the parameters of these  diffusion processes (\cite{Donnet2009}, \cite{Delattre2013}, \cite{Donnet2013}, \cite{Delattre2018}). To our knowledge, the framework of mixed-effects models has rarely been used to analyse epidemic data, except in a very few studies. Among these, in (\cite{prague:hal-02555100}), the dynamics of the first epidemic wave of COVID-19 in France were analysed using an ODE system incorporating random parameters to take into account the variability of the dynamics between regions. Using a slightly different approach to tackle data from multiple epidemics, \cite{Breto2020} proposed a likelihood-based inference method using particle filtering techniques for non-linear and partially observed models. In particular, these models incorporate unit-specific parameters and shared parameters.  \\

\noindent In addition to the specific problem of variability reflected in multiple data sets, observations of epidemic dynamics are often incomplete in various ways: only certain health states are observed (e.g. infected individuals), data are temporally discretized or aggregated, and subject to observation errors (e.g. under-reporting, diagnosis errors). Because of this incompleteness together with the non-linear structure of the epidemic models, the computation of the maximum likelihood estimator (MLE) is often not explicit. In hidden or latent variable models which are appropriate representations of incompletely observed epidemic dynamics, estimation techniques based on Expectation-Maximization (EM) algorithm can be implemented in order to compute the MLE (see e.g. \cite{Dempster1977}). However, the E-step of the EM algorithm  requires that, for each parameter value $\theta$, the conditional expectation of the complete log-likelihood given the observed data, $\mathcal{Q}(\theta)$, can be computed. In mixed-effects models, there is generally no closed form expression for $\mathcal{Q}(\theta)$. In such cases, this quantity can be approximated using a Monte-Carlo procedure (MCEM, \cite{Wei1990}), which is computationally very demanding. A more efficient alternative is the SAEM algorithm (\cite{Delyon1999}), often used in the framework of mixed-effects models (\cite{Kuhn2005}), which combines at each iteration the simulation of unobserved data under the conditional distribution given the observations and a stochastic approximation procedure of $\mathcal{Q}(\theta)$ (see also \cite{Delattre2013}, \cite{Donnet2014} for the study and implementation of the SAEM algorithm for mixed-effects diffusion models). \\

\noindent In this paper, focusing on the inference for multiple epidemic dynamics, we intend to meet two objectives. The first objective is to propose a finer modeling of multiple epidemics through a unique mixed-effects model, incorporating a stochastic representation of each epidemic. The second objective is to develop an appropriate method for jointly estimating model parameters from noisy and partial observations, able to estimate rigorously and explicitly the inter-epidemic variability. Thus, the main expected contribution is to provide accurate estimates of common and epidemic-specific parameters and to provide elements for the interpretation of the mechanisms underlying the variability between epidemics of the same nature occurring in different locations or over distinct time periods.
For this purpose, we extend the Gaussian state-space model introduced in (\cite{Narci2020}) for single epidemics to a model with mixed effects on the parameters describing simultaneously several epidemics and their observations. Then, following (\cite{Delattre2013}) and building on the Kalman filtering-based inference method proposed in (\cite{Narci2020}), we propose to couple the SAEM algorithm with Kalman-like filtering to estimate model parameters. The performances of the estimation method are investigated on simulations mimicking noisy prevalence data (\textit{i.e.} the number of cases of disease in the population at a given time or over a given period of time). The method is then applied to the case of influenza epidemics in France over several years using noisy incidence data (\textit{i.e.} the number of newly detected cases of the disease at a given time or over a given period of time), by proposing a new version of the filtering algorithm to handle this type of data. \\

\noindent The paper is organized as follows. In Section \ref{sec:model} we describe
the epidemic model for a single epidemic, specified for both prevalence and incidence data, and its extension to account for several epidemics through a two-level representation using the framework of mixed-effects models. Section \ref{sec:inference} contains the maximum likelihood estimation method and convergence results of the SAEM algorithm. In Section \ref{sec:estimsimu}, the performances of our inference method are assessed on simulated noisy prevalence data generated by SIR epidemic dynamics sampled at discrete time points. Section \ref{sec:estimrealdata} is dedicated to the application case, the influenza outbreaks in France from 1990 to 2017.  Section \ref{sec:discussion} contains a discussion and concluding remarks.

\section{A mixed-effects approach for a state-space epidemic model for multiple epidemics}\label{sec:model} 

First, we sum up the approach developed in (\cite{Narci2020}) in the case of single epidemics for prevalence data and extend it to incidence data (Section \ref{single_model}). By extending this approach, we propose a model for simultaneously considering several epidemics, in the framework of mixed-effects models (Section \ref{multiple_epid_mod}).

\subsection {The basics of the modeling framework for the case of a single epidemic}
\label{single_model}

\paragraph{The epidemic model} \label{dynmod}
Consider an epidemic in a closed population of size $N$ with homogeneous mixing, whose dynamics are represented by a stochastic compartmental model with $d+1$ compartments corresponding to the successive health states of the infectious process within the population. These dynamics are described by a density-dependent Markov jump process $\mathcal{Z}(t)$ with state space $\{0,\dots, N\}^d$ and transition rates depending on a multidimensional parameter $\zeta$.
Assuming that $\mathcal{Z}(0)/N \rightarrow x_0 \neq (0,\dots,0)'$, the normalized process $\mathcal{Z}(t)/N$  representing the respective proportions of population in each health state converges, as $N \rightarrow \infty$, to a classical and well-characterized ODE:
\begin{equation} \label{ODEgen}
\frac{\partial x}{\partial t}(\zeta,t)= b( \eta, x(\zeta,t)); \quad x(0)=x_0,
\end{equation}
where $\eta=(\zeta,x_0)$ and $b(\eta,\cdot)$ is explicit and easy to derive from the Q-matrix of process $\mathcal{Z}(t)$ (see (\cite{Guy2015}), (\cite{Narci2020})).\\

\noindent Two stochastic approximations of $\mathcal{Z}(t)/N$ are available: a $d$-dimensional diffusion process $Z(t_k)$ with drift coefficient $b(\eta,\cdot)$ and diffusion matrix $\frac{1}{N} \Sigma( \eta,\cdot)$ (which is also easily deducible from the jump functions of the density-dependent jump process, see e.g. (\cite{Narci2020})), and a time-dependent Gaussian process $G_N(t)$ with small variance coefficient (see e.g. \cite{Pardoux2020}), having for expression
\begin{equation}\label{Xeta}
G_N(t)= x(\eta,t) + \frac{1}{\sqrt{N}} g(\eta,t),
\end{equation}
where $g(\eta,t)$ is a centered  Gaussian process with explicit covariance matrix. There is a link between these two processes: let
$W(t)$ be a Brownian motion in $\mathbb{R}^d$, then
$g(\eta,t)$ is the centered Gaussian process  
$$g(\eta,t) = \int_0^t \Phi(\eta, t,u)\sigma(\eta, x(\eta,u)) dW(u), \quad \mbox{where } \sigma(\eta,x)\sigma(\eta,x)^{\prime}= \Sigma(\eta,x), $$
 and $\Phi(\eta, t,s) $ is the 
$d\times d$ resolvent matrix associated to \eqref{ODEgen}
\begin{equation}\label{dPhi}
\Phi(\eta, t,s) = \exp\left(\int_{s}^{t} \nabla_x b(\eta,x(\eta,u)) \ du\right),
\end{equation}
with $\nabla_x b (\eta, x)$ denoting the matrix $( \frac{\partial b_i}{\partial x_j} (\eta, x))_{1\leq i,j\leq d}$. In the sequel, we rely on the Gaussian process (\ref{Xeta}) to represent epidemic dynamics. \\

\noindent The epidemic is observed at discrete times $t_0=0 < t_1 ,\cdots,<t_n=T$, where $n$ is the number of observations. Let us assume that the observation times $t_k$ are regularly spaced, that is $t_k = k \Delta$ with $\Delta$ the time step (but the following can be easily adapted to irregularly spaced observation times). Setting $X_k:= G_N(t_k)$ and $X_0= x_0$, the model can be written under the auto-regressive AR(1) form
\begin{equation}\label{ARModel}
X_{k} = F_k(\eta) + A_{k-1} (\eta)X_{k-1} + V_{k}, \quad \mbox{with  } V_{k} \sim  \mathcal{N}_d\left(0, T_k(\eta,\Delta)\right) \text{ and } k\geq 1.
\end{equation}
All the quantities in \eqref{ARModel} have explicit expressions with respect to the parameters. 
Indeed, using \eqref{ODEgen} and \eqref{dPhi}, we have
\begin{align}
A_{k-1}(\eta) & =  A(\eta,t_{k-1}) =  \Phi(\eta,t_k, t_{k-1}), \label{model_quantities1}\\
F_k(\eta) &=  F(\eta,t_k)  =  x(\eta,t_k)-  \Phi(\eta,t_k, t_{k-1})x(\eta,t_{k-1}), \label{model_quantities2} \\
T_k(\eta,\Delta)&=  \frac{1}{N}\int_{t_{k-1}}^{t_k} \Phi(\eta,t_k,s) \Sigma(\eta,x(\eta,s))\; \Phi^t(\eta,t_k,s)ds. \label{model_quantities3}
\end{align}

%

\begin{figure}[h!] 
\centering
\begin{tikzpicture}[node distance=3cm,auto,>=latex']
    \node[int] (c) [] {$S$};

    \node [int] (d) [right of=c, node distance=4cm] {$I$};
    
    \node [int] (e) [right of=d, node distance=4cm] {$R$};

    \draw[->, ultra thick, black] (c) edge node {$\lambda I/N$} (d);
    \draw[->, ultra thick, black] (d) edge node {$\gamma$} (e);
\end{tikzpicture}
\caption{SIR compartmental model with three blocks corresponding respectively to susceptible (S), infectious (I) and recovered (R) individuals. Transitions of individuals
from one health state to another are governed by the transmission rate $\lambda$ and the recovery rate $\gamma$, respectively.}
\label{fig:SIR_rates}
\end{figure}
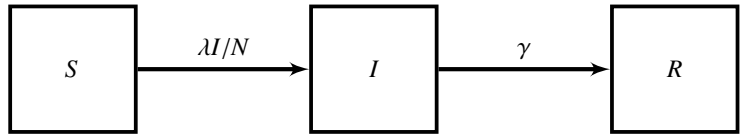

\noindent \textbf{Example: SIR  model.} 
As an illustrative example, we use the simple
SIR epidemic model described in Figure \ref{fig:SIR_rates}, but other models can be considered (see e.g. the SEIR model, used in Section \ref{sec:estimrealdata}).\\

\noindent In the SIR model, $d=2$ and $\mathcal{Z}(t)=\left(S(t),I(t)\right)'$. The parameters involved in the transition rates are $\lambda$ and $\gamma$ and the initial proportions of susceptible and infectious individuals are $x_0= (s_0,i_0)'$. Denoting $\eta=(\lambda,\gamma,s_0,i_0)'$, the ODE satisfied by $x(\eta,t)= (s(\eta,t), i(\eta,t))'$ is
\begin{align}
\label{SIRODE}
\begin{cases}
\frac{\partial s}{\partial t}(\eta,t)= -\lambda s(\eta,t) i(\eta,t);  \quad  s(\eta,0)= s_0,\\
\frac{\partial i}{\partial t}(\eta,t)= \lambda s(\eta,t) i(\eta,t)- \gamma i(\eta,t);\quad    \quad i(\eta,0)= i_0.
\end{cases}
\end{align}
When there is no ambiguity, we denote by $s$ and $i$ the solution of  \eqref{SIRODE}.
Then, the functions
$b(\eta,\cdot)$, $\Sigma(\eta,\cdot)$ and $\sigma(\eta,\cdot)$ are
\begin{equation*}
\label{derive}
 b(\eta, s,i) = \begin{pmatrix} - \lambda s i \\ \lambda s i -\gamma i  \end{pmatrix}; \quad  \Sigma(\eta,s,i) = \begin{pmatrix} \lambda s i & - \lambda s i \\ - \lambda s i & \lambda s i + \gamma i \end{pmatrix},\;\sigma(\eta,s,i) = \begin{pmatrix} \sqrt{\lambda s i} & 0 \\ - \sqrt{\lambda s i} &  \sqrt{\gamma i} \end{pmatrix}.
\end{equation*}

\noindent We refer the reader to Appendix A for the computation of $b(\eta,\cdot)$, $\Sigma(\eta,\cdot)$ and $\sigma(\eta,\cdot)$ in the SEIR model. Another parameterization, involving the basic reproduction number $R_0=\frac{\lambda}{\gamma}$ and the infectious period $d=\frac{1}{\gamma}$, is more often used for SIR models. Hence, we set $\eta=(R_0,d,s_0,i_0)'$.


%
%
\paragraph{Observation model for prevalence data}
Following (\cite{Narci2020}),
we assume that observations are made at times $t_k=k\Delta, k=1,\dots,n$, and that some health states are not observed. The dynamics is described by the $d$-dimensional $AR(1)$ model detailed in \eqref{ARModel}. Some coordinates are not observed and various sources of noise systematically affect the observed coordinates (measurement errors, observation noises, under-reporting, etc.). This is taken into account by introducing an additional parameter $\mu$, governing both the levels of noise and the amount of information which is available from the $q \leq d$ observed coordinates, and an operator $B(\mu): \mathbb{R}^d \rightarrow \mathbb{R}^q$.
Moreover, we assume that, conditionally on the random variables $(B(\mu)X_k, k=1, \dots,n)$, these noises are independent but not identically distributed. 
We approximate their distributions by $q$-dimensional Gaussian distributions with covariance matrix $P_k(\eta,\mu)$ depending on $\eta$ and $\mu$. This yields that the observations $(Y_k)$ satisfy
\begin{equation}\label{Yk}
Y_{k}= B(\mu)X_{k}+W_{k}, \mbox{ with } W_k  \sim {\cal N}_q(0,P_k( \eta,\mu)).
\end{equation}
Let us define a global parameter describing both the epidemic process and the observational process,  
\begin{equation}\label{phi}
\phi= (\eta,\mu).
\end{equation}

\noindent Finally, joining \eqref{ARModel}, \eqref{Yk} and \eqref{phi} yields the formulation (for both epidemic dynamics and observation process) required to implement Kalman filtering methods in order to estimate the epidemic parameters: 
\begin{align}
\label{KalmanModel}
\begin{cases}
X_{k} = F_k(\eta) + A_{k-1} (\eta)X_{k-1} + V_{k},\;\mbox{ with } V_{k} \sim  \mathcal{N}_d\left(0, T_k(\eta,\Delta)\right), \ k\geq 1, \\ 
Y_{k}= B(\mu)X_{k}+W_{k},\mbox{ with } W_{k} \sim  \mathcal{N}_q\left(0, P_k(\phi)\right).
\end{cases}
\end{align}

\noindent \textbf{Example: SIR  model (continued).} 
The available observations could be noisy proportions of the number of infectious individuals at discrete times $t_k$.
Denoting by $p$ the reporting rate, one could define the operator $B(\mu) = B(p) = (0 \ \ \ p)$ and the covariance error as $P_k(\phi)=\frac{1}{N}p(1-p)i(\eta,t_k)$ with $i(\eta,t)$ satisfying \eqref{SIRODE}. The expression of $P_k(\phi)$ mimics the variance that would arise from assuming the observations to be obtained as binomial draws of the infectious individuals.

\paragraph{Observation model for incidence data} 
For this purpose, we have extended the framework developed in (\cite{Narci2020}). For some compartmental models, the observations (incidence) at times $t_k$ can be written as the increments of a single or more coordinates, that is $\tilde{B}(\mu)(X_{k-1}-X_k)$ where, as above, $\tilde{B}(\mu) : \mathbb{R}^d \rightarrow \mathbb{R}^q$ is a given operator and $\mu$ are emission parameters. Let us write the epidemic model in this framework. For $k=1,\ldots,n$, let 
\begin{equation*}
 \Delta_k X = X_k - X_{k-1}. 
\end{equation*} 

\noindent From (\ref{KalmanModel}), the following holds, denoting by $I_d$ the $d \times d$ identity matrix,
\begin{equation}
\label{eqIncrements}
\Delta_k X =  F_k(\eta) + (A_{k-1}(\eta)-I_d) X_{k-1} + V_k.
\end{equation}
\noindent As $X_{k-1} = \sum_{l=1}^{k-1} \Delta_l X + x_0$, (\ref{eqIncrements}) becomes: 
 \begin{equation}\label{Deltak}
  \Delta_k X =  G_k(\eta) + (A_{k-1}(\eta)-I_d) \sum_{l=1}^{k-1} \Delta_l X + V_k, \mbox{ with }
\end{equation}
\begin{equation}\label{Gk}
 G_k(\eta) = x(\eta,t_k)-x_0 -  \Phi(\eta, t_k, t_{k-1})(x(\eta,t_{k-1})-x_0).
\end{equation}

\noindent To model the errors that affect the data collected $(Y_k)$, we assume that, conditionally on $(\Delta_k X, k=1,\dots,n)$, the observations are independent and proceed to the same approximation for their distributions
\begin{equation}\label{incidence}
 Y_k = \tilde{B}(\mu) \Delta_k X + \tilde{W}_k ; \quad \mbox{with  } \tilde{W}_k \sim {\cal N}_q(0, \tilde{P}_k(\phi)).
\end{equation}
Consequently, using \eqref{Deltak}, \eqref{Gk} and \eqref{incidence}, the epidemic model for incidence data is adapted as follows: 
\begin{align}    
\label{KalmanModel_inc}
\begin{cases}
\Delta_k X = G_k(\eta) + (A_{k-1}(\eta)-I_d) \sum_{l=1}^{k-1} \Delta_l X + V_k, \\
 Y_k = \tilde{B}(\mu) \Delta_k X + \tilde{W}_k.
\end{cases}
\end{align}
Contrary to \eqref{ARModel}, $(\Delta_k X, k=1,\dots,n)$ is not Markovian since it depends on all the past observations. Therefore, it does not possess the required properties of classical Kalman filtering methods. We prove in Appendix B that we can propose an iterative procedure and define a new filter to compute recursively the conditional distributions describing the updating and prediction steps together with the marginal distributions of the observations from the model \eqref{KalmanModel_inc}. \\
 
\noindent \textbf{Example: SIR model (continued).} Here, $\Delta_k X=\left(\frac{\Delta_k S}{N},\frac{\Delta_k I}{N}\right)'$ and the number of new infectious individuals at times $t_k$  is given by $ \int_{t_{k-1}}^{t_k} \lambda S(t) \frac{I(t)}{N} \ dt = - \Delta_k S$. Observing a proportion $p$ of the new infectious individuals would lead to the operator $\tilde{B}(\mu) = B(p) = (-p \ \ \ 0)$. Mimicking binomial draws, the covariance error could be chosen as $\tilde{P}_k(\phi)=\frac{1}{N}p(1-p)(s(\eta,t_{k-1})-s(\eta,t_k))$ where $s(\eta,t)$ satisfies \eqref{SIRODE}.

\subsection{Modeling framework for multiple epidemics} 
\label{multiple_epid_mod}
Consider now the situation where a same outbreak occurs in many regions or at different periods simultaneously. We use the index $1\le u \le U$ to describe the quantities for each unit (e.g. region or period), where $U$ is the total number of units.
Following Section \ref{single_model}, for unit $u$, the epidemic dynamics are represented by the $d$-dimensional process $(X_u(t))_{t\ge 0}$ corresponding to $d+1$ infectious states (or compartments) with 
state space $E=[0,1]^d$. It is assumed that $(X_u(t))_{t\ge 0}$ is observed at discrete times $t_k=k \Delta$ on $[0,T_u]$, $T_u=n_u \Delta$, where $\Delta$ is a fixed time step and 
$n_u$ is the number of observations, and that $Y_{u,k}$ are the observations at times $t_k$. Each of these dynamics has its own epidemic and observation parameters, denoted $\phi_u$.  \\
 
\noindent To account for intra- and inter-epidemic variability, a two level representation is considered, in the framework of mixed-effects models. 
First, using the discrete-time Gaussian state-space for prevalence (\ref{KalmanModel}) or for incidence data (\ref{KalmanModel_inc}), the intra-epidemic variability is described. Second, the inter-epidemic variability is characterized by specifying a set of random parameters for each epidemic.

\paragraph{1. Intra-epidemic variability}
Let us define $X_{u,k}:= X_u(t_k)$, $X_{u,0}= x_{u,0}$ and $\Delta_k X_u := X_u(t_{k}) - X_u(t_{k-1})$. 
Using \eqref{phi}, conditionally to $\phi_u=\varphi$, the epidemic observations for unit $u$ are  described as in Section \ref{single_model}. \\ 

\noindent For  prevalence data, $ 1\leq k\leq n_u$,
\begin{align}\label{Kalman_gen}
\begin{cases}
X_{u,k}= F_k(\varphi) + A_{k-1} (\varphi)X_{u,k-1} + V_{u,k},\quad \mbox{with } V_{u,k} \sim  \mathcal{N}_d\left(0, T_k(\varphi,\Delta)\right), \\ 
Y_{u,k}= B(\varphi)X_{u,k}+W_{u,k}, \quad \mbox{ with } W_{u,k} \sim {\cal N} _q( 0, P_k(\varphi)),
\end{cases}
\end{align}
(see (\ref{model_quantities1}), (\ref{model_quantities2}) and (\ref{model_quantities3}) for the expressions of  $F_k(\cdot)$, $A_{k-1}(\cdot)$, $T_k(\cdot) $ and \eqref{Yk} for $B(\cdot)$ and $P_k(\cdot)$). \\

\noindent For incidence data, 
\begin{align}
\label{Kalman_gen_inc}
\begin{cases}
\Delta_k X_u = G_k(\varphi) + (A_{k-1}(\varphi)-I_d) \sum_{l=1}^{k-1} \Delta_l X_u + V_{u,k},\\
Y_{u,k}= \tilde{B}(\varphi) \Delta_k X_u + \tilde{W}_{u,k}  \quad \mbox{ with } \tilde{W}_{u,k} \sim {\cal N} _q( 0, \tilde{P}_k(\varphi)),
\end{cases}
\end{align}
(see (\ref{Gk}) for the expression of $G_k(\cdot)$ and \eqref{incidence} for $\tilde{B}(\cdot)$ and $\tilde{P}_k(\cdot)$).

\paragraph{2. Inter-epidemic variability} 

We assume that the epidemic-specific parameters $(\phi_u,1\le u \le U)$
are independent and identically distributed (i.i.d) random variables with distribution defined as follows,
\begin{align} \label{parms_indiv}
\begin{cases}
  \phi_u &= h(\beta,\xi_u), \\
  \xi_u &\sim \mathcal{N}_c(0,\Gamma),
  \end{cases}
\end{align}
where $c=\text{dim} \ (\phi_u)$ and $h(\beta,x): \mathbb{R}^{c} \times   \mathbb{R}^{c}\rightarrow  \mathbb{R}^{c}$.
The vector   
$h(\beta,x)=\left(h_1(\beta,x),\ldots,h_c(\beta,x)\right)'$ contains known link functions (a classical way to obtain parameterizations easier to handle), $\beta \in \mathbb{R}^{c}$ is a vector of fixed effects and $\xi_1,\ldots,\xi_U$ are random effects modeled by $U$ i.i.d centered random variables. The fixed and random effects respectively describe the average general trend shared by all epidemics and the differences between epidemics. Note that it is sometimes possible to propose a more refined description of the inter-epidemic variability by including unit-specific covariates in \eqref{parms_indiv}. This is not considered here, without loss of generality.\\

\noindent \textbf{Example: SIR model (continued).} Let $s_{0,u} = \frac{S_u(0)}{N_u}$ and $i_{0,u} = \frac{I_u(0)}{N_u}$ where $N_u$ is the population size in unit $u$. The random parameter is  $\phi_u = (R_{0,u},d_u,p_u,s_{0,u},i_{0,u})'$ and has to fulfill the constraints 
$$R_{0,u} > 1;\; d_u > 0; \;  0< p_u < 1; \; 0< s_{0,u}, i_{0,u} < 1,\;  s_{0,u}+i_{0,u} \le 1.$$

\noindent To meet these constraints, one could introduce the following function $h(\beta,x):  \mathbb{R}^{5} \times   \mathbb{R}^{5}\rightarrow  \mathbb{R}^{5}$:
\begin{align}\label{transfo}
\begin{cases}
  h_1(\beta,\xi_u) &= \exp\left[\beta_1+\xi_{1,u}\right]+1, \\
 h_2(\beta,\xi_u) &= \exp\left[\beta_2+\xi_{2,u}\right], \\
h_3(\beta,\xi_u) &= \frac{1}{1+\exp\left[-(\beta_3+\xi_{3,u})\right]}, \\
h_4(\beta,\xi_u) &= \frac{1}{1+\exp\left[-(\beta_4+\xi_{4,u})\right]+\exp\left[-(\beta_5+\xi_{5,u})\right]}, \\ 
h_5(\beta,\xi_u) &= \frac{\exp\left[-(\beta_4+\xi_{4,u})\right]}{1+\exp\left[-(\beta_4+\xi_{4,u})\right]+\exp\left[-(\beta_5+\xi_{5,u})\right]},
\end{cases}
\end{align}
where $\xi_u \sim_{i.i.d.} \mathcal{N}_5(0,\Gamma)$ and $\phi_u= h(\beta,\xi_u)$. \\

\noindent In this example, we supposed that all the parameters have both fixed and random effects, but it is also possible to consider a combination of random-effect parameters and purely fixed-effect parameters (see Section \ref{sec:datasimulations} for instance). 
 
\section{Parametric inference}
\label{sec:inference}

To estimate the model parameters $\theta=(\beta,\Gamma)$, with $\beta$ and $\Gamma$ defined in \eqref{parms_indiv}, containing the parameters modeling the intra- and inter-epidemic variability, we develop an algorithm in the spirit of (\cite{Delattre2013}) allowing to derive the maximum likelihood estimator (MLE). 

\subsection{Maximum likelihood estimation} 

The model introduced in Section \ref{multiple_epid_mod} can be seen as a latent variable model with $\mathbf{y} = (y_{u,k}, 1 \le u \le U, 0\le k \le n_u)$ the observed data and $\mathbf{\Phi} = (\phi_u,1\le u \le U)$
the latent variables. Denote respectively by $p(\mathbf{y};\theta$), $p(\mathbf{\Phi};\theta)$ and $p(\mathbf{y}|\mathbf{\Phi};\theta)$  
the probability density of the observed data, of the random effects and of the observed data given the unobserved ones. 
By independence of the $U$ epidemics, the likelihood of the observations $\mathbf{y}_u=(y_{u,1},\ldots,y_{u,n_u})$ is given by: 
\begin{equation*}\label{p_obs}
 p(\mathbf{y};\theta) = \prod_{u=1}^{U} p(\mathbf{y}_u;\theta).
\end{equation*}
Computing the distribution $p(\mathbf{y}_u;\theta)$ of the observations for any epidemic $u$ requires the integration of the conditional density of the data given the unknown random effects $\phi_u$ with respect to the density 
of the random parameters:
\begin{equation} 
\label{p_obs_u}
 p(\mathbf{y}_u;\theta) = \int p(\mathbf{y}_u|\mathbf{\phi}_u;\theta) p(\mathbf{\phi}_u;\theta) \ d\mathbf{\phi}_u.
\end{equation}

\noindent Due to the non-linear structure of the proposed model, the 
 integral in (\ref{p_obs_u}) is not explicit. Moreover, the computation of $p(\mathbf{y}_u|\mathbf{\phi}_u;\theta)$ is not straightforward due to the presence of latent states in the model. Therefore, the inference algorithm needs to account for these specific features. \\

\noindent Let us first deal with the integration with respect to the unobserved random variables $\mathbf{\phi}_u$. In latent variable models, the use of the EM algorithm (\cite{Dempster1977}) allows to compute iteratively the MLE. Iteration $k$ of the EM algorithm combines two steps: (1) the computation of 
the conditional expectation of the complete log-likelihood given the observed data and the current parameter estimate $\theta_k$, denoted $\mathcal{Q}(\theta|\theta_k)$ (E-step); (2) the update of the parameter estimates by maximization of $\mathcal{Q}(\theta|\theta_k)$ (M-step). 
In our case, the E-step cannot be performed because $\mathcal{Q}(\theta|\theta_k)$ does not have a simple analytic expression. We rather implement a Stochastic Approximation-EM (SAEM, \cite{Delyon1999}) which combines at 
each iteration the simulation of unobserved data under the conditional distribution given the observations (S-step) and a stochastic approximation of $\mathcal{Q}(\theta|\theta_k)$ (SA-step).

\paragraph{a) General description of the SAEM algorithm} Given some initial value $\theta_{0}$, iteration $m$ of the SAEM algorithm consists in the three following steps: 

\begin{itemize}
 \item[] (S-step) Simulate a realization of the random parameters $\mathbf{\Phi}_m$ under the conditional distribution given the observations for a current parameter $\theta_{m-1}$ denoted $p(\cdot|\mathbf{y};\theta_{m-1})$.
 \item[] (SA-step) Update $\mathcal{Q}_m(\theta)$ according to
 \begin{equation*}
  \mathcal{Q}_m(\theta) = \mathcal{Q}_{m-1}(\theta) + \alpha_m(\log p(\mathbf{y},\mathbf{\Phi}_m;\theta) - \mathcal{Q}_{m-1}(\theta)),
 \end{equation*}
where $(\alpha_m)_{m \ge 1}$ is a sequence of positive step-sizes s.t. $\sum_{m=1}^{\infty} \alpha_m = \infty$ and $\sum_{m=1}^{\infty} \alpha_m^2 < \infty$. \\
 \item[] (M-step) Update the parameter estimate by maximizing $\mathcal{Q}_m(\theta)$ 
 \begin{equation*}
  \theta_m = \text{arg max}_{\theta} \ \mathcal{Q}_m(\theta).
 \end{equation*}

\end{itemize}

\noindent In our case, an exact sampling under $p(\cdot|\mathbf{y};\theta_{m-1})$ in the S-step is not feasible.
In such intractable cases, MCMC algorithms such as Metropolis-Hastings algorithm can be used (\cite{Kuhn2004}).

\paragraph{b) Computation of the S-step by combining the Metropoligs-Hastings algorithm with Kalman filtering techniques}

\noindent In the sequel,
we combine the S-step of the SAEM algorithm with a MCMC procedure.\\ 

\noindent For a given parameter value $\theta$, a single iteration of the Metropolis-Hastings algorithm consists in:

\begin{itemize}
 \item[(1)] Generate a candidate $\mathbf{\Phi}^{(c)} \sim q(\cdot|\mathbf{\Phi}_{m-1},\mathbf{y};\theta)$ for a given proposal distribution $q$ 
 \item[(2)] Take
 \begin{equation*}
  \mathbf{\Phi}_{m} = \begin{cases}
                      \mathbf{\Phi}_{m-1} \text{ with probability } 1 - \rho(\mathbf{\Phi}_{m-1},\mathbf{\Phi}^{(c)}),\\
                      \mathbf{\Phi}^{(c)} \text{ with probability } \rho(\mathbf{\Phi}_{m-1},\mathbf{\Phi}^{(c)}),
                     \end{cases}
 \end{equation*}
 where \begin{equation}
 \label{rate_MH}
        \rho(\mathbf{\Phi}_{m-1},\mathbf{\Phi}^{(c)}) = \text{min}\left[1, \frac{p(\mathbf{y}|\mathbf{\Phi}^{(c)}; \theta) \
       p(\mathbf{\Phi}^{(c)}; \theta) \ q(\mathbf{\Phi}_{m-1}|\mathbf{\Phi}^{(c)},\mathbf{y}; \theta)}{p(\mathbf{y}|\mathbf{\Phi}_{m-1}; \theta) \ p(\mathbf{\Phi}_{m-1}; \theta) \
        q(\mathbf{\Phi}^{(c)}|\mathbf{\Phi}_{m-1},\mathbf{y}; \theta)}\right].
       \end{equation}

\end{itemize}





\noindent To compute the rate of acceptation of the Metropolis-Hastings algorithm in \eqref{rate_MH}, we need to calculate
\begin{equation*}
\label{p_obs_cond}
 p(\mathbf{y}_u|\mathbf{\phi}_u;\theta) = p(y_{u,0}|\phi_u;\theta) \prod_{k=1}^{n_u} p(y_{u,k}|y_{u,0},\ldots,y_{u,k-1},\mathbf{\phi}_u;\theta), \ 1 \le u \le U.
\end{equation*}

\noindent Let $y_{u,k:0}:=(y_{u,0},\ldots,y_{u,k})$, $k\geq 1$. In both models (\ref{Kalman_gen}) and (\ref{Kalman_gen_inc}), the conditional densities $p(y_{u,k}|y_{u,k-1:0},\mathbf{\phi}_u;\theta)$ are Gaussian densities. In model (\ref{Kalman_gen}) involving prevalence data, their means and variances can be exactly computed with Kalman filtering techniques (see (\cite{Narci2020})). In model (\ref{Kalman_gen_inc}), the Kalman filter can not be used in its standard form. We therefore develop an alternative filtering algorithm.   \\



\noindent From now on, we omit the dependence in $u$ and $\mathbf{\Phi}$ for sake of simplicity. 


\paragraph{Prevalence data}

Let us consider model (\ref{KalmanModel}) and recall the successive steps of the filtering developed in (\cite{Narci2020}).
 Assume that $X_0 \sim {\cal N}_d(x_0, T_0)$ and set  ${\hat X}_0=  x_0,{\hat \Xi}_0= T_0$.
Then, the Kalman filter consists in recursively computing for $k\ge 1$: 
\begin{itemize}
\item[1.] Prediction: $\mathcal{L}(X_{k+1}|Y_k,\ldots,Y_1) = \mathcal{N}_d(\widehat{X}_{k+1},\widehat{\Xi}_{k+1})$
\begin{align*}
 \widehat{X}_{k+1} &= F_{k+1} + A_k \overline{X}_k \\
 \widehat{\Xi}_{k+1} &= A_k \overline{T}_k A_k' + T_{k+1}
\end{align*}
\item[2.] Updating: $\mathcal{L}(X_k|Y_k,\ldots,Y_1) = \mathcal{N}_d(\overline{X}_k,\overline{T}_k)$
 \begin{align*}
  \overline{X_k} &= \widehat{X}_k + \widehat{\Xi}_k B' (B \widehat{\Xi}_k B' + P_k)^{-1}(Y_k - B\widehat{X}_k) \\
  \overline{T_k} &= \widehat{\Xi}_k - \widehat{\Xi}_k B' (B \widehat{\Xi}_k B' + P_k)^{-1} B \widehat{\Xi}_k
 \end{align*}
\item[3.] Marginal: $\mathcal{L}(Y_{k+1}|Y_k,\ldots,Y_1) = \mathcal{N}(\widehat{M}_{k+1},\widehat{\Omega}_{k+1})$
\begin{align*}
 \widehat{M}_{k+1} &= B \widehat{X}_{k+1} \\
 \widehat{\Omega}_{k+1} &= B \widehat{\Xi}_{k+1} B' + P_{k+1}
\end{align*}
\end{itemize}

\paragraph{Incidence data} Let us consider model (\ref{KalmanModel_inc}). Assume that
 $\mathcal{L}(\Delta_1 X) = \mathcal{N}_d(G_1,T_1)$ and $\mathcal{L}(Y_1 | \Delta_1 X) = \mathcal{N}_q(\tilde{B}\Delta_1 X, \tilde{P}_1)$. 
Let $\widehat{\Delta_1 X} = G_1 = x(t_1) - x_0$ and $\widehat{\Xi}_1 = T_1$. Then, at iterations $k\ge 1$, the filtering steps are: \\

\begin{itemize}
\item[1.] Prediction: $\mathcal{L}(\Delta_{k+1} X|Y_k,\ldots,Y_1) = \mathcal{N}_d(\widehat{\Delta_{k+1} X},\widehat{\Xi}_{k+1})$
\begin{align*}
 \widehat{\Delta_{k+1} X} &= G_{k+1} + (A_k-I_d) \left(\sum_{l=1}^{k}\overline{\Delta_l X}\right) \\
 \widehat{\Xi}_{k+1} &= (A_k-I_d) \left(\sum_{l=1}^{k}\overline{T}_l\right) (A_k-I_d)' + T_{k+1}
\end{align*}
 \item[2.] Updating: $\mathcal{L}(\Delta_k X|Y_k,\ldots,Y_1) = \mathcal{N}_d(\overline{\Delta_k X},\overline{T}_k)$
 \begin{align*}
  \overline{\Delta_k X} &= \widehat{\Delta_k X} + \widehat{\Xi}_k \tilde{B}' (\tilde{B} \widehat{\Xi}_k \tilde{B}' + \tilde{P}_k)^{-1}(Y_k - \tilde{B}\widehat{\Delta_k X}) \\
  \overline{T}_k &= \widehat{\Xi}_k - \widehat{\Xi}_k \tilde{B}' (\tilde{B} \widehat{\Xi}_k \tilde{B}' + \tilde{P}_k)^{-1} \tilde{B} \widehat{\Xi}_k
 \end{align*}
\item[3.] Marginal: $\mathcal{L}(Y_{k+1}|Y_k,\ldots,Y_1) = \mathcal{N}(\widehat{M}_{k+1},\widehat{\Omega}_{k+1})$
\begin{align*}
 \widehat{M}_{k+1} &= \tilde{B} \widehat{\Delta_{k+1} X} \\
 \widehat{\Omega}_{k+1} &= \tilde{B} \widehat{\Xi}_{k+1} \tilde{B}' + \tilde{P}_{k+1}
\end{align*}
\end{itemize}

\noindent The equations are deduced in Appendix B, the difficult point lying in the prediction step, \textit{i.e.} the derivation of  the conditional distribution ${\cal  L}( \Delta_{k+1} X|Y_{k},\cdots, Y_1)  $.

\subsection{Convergence of the SAEM-MCMC algorithm}

Generic assumptions guaranteeing the convergence of the SAEM-MCMC algorithm were stated in (\cite{Kuhn2004}). These assumptions mainly concern the regularity of the model (see assumptions \textbf{(M1-M5)}) and the properties of the MCMC procedure used in step S (\textbf{SAEM3'}). Under these assumptions, and providing that the step sizes $(\alpha_m)$ are such that $\sum_{m=1}^{\infty} \alpha_m = \infty$ and $\sum_{m=1}^{\infty} \alpha_m^2 < \infty$, then the sequence $(\theta_m)$ obtained through the iterations of the SAEM-MCMC algorithm converges almost surely toward a stationary point of the observed likelihood. \\

\noindent Let us remark that by specifying the inter-epidemic variability through the modeling framework of Section \ref{multiple_epid_mod}, our approach for multiple epidemics fulfills the exponentiality condition stated in \textbf{(M1)} provided that all the components of $\phi_u$ are random. 
Hence the algorithm proposed above converges almost surely toward a stationary point of the observed likelihood under the standard regularity conditions stated in \textbf{(M2-M5)} and assumption \textbf{(SAEM3')}.

 \section{Assessment of parameter estimators performances on simulated data}
\label{sec:estimsimu}

First, the performances of our inference method are assessed on simulated stochastic SIR dynamics. Second, the estimation results are compared with those obtained by an empirical two-step approach. \\

\noindent For a given population of size $N$ and given parameter values, we use the Gillespie algorithm (\cite{Gillespie1977}) to simulate a two-dimensional Markov jump process  ${\cal Z}(t)= (S(t),I(t))'$. Then, choosing a sampling interval $\Delta$ and a reporting rate $p$, we consider prevalence data $(O(t_k), k=1,\ldots,n)$ simulated as binomial trials from a single coordinate of the system $I(t_k)$.

\subsection{Simulation setting}
\label{sec:datasimulations}

\paragraph{Model} Recall that the epidemic-specific parameters are $\phi_u = \left(R_{0,u}, d_u, p_u,s_{0,u},i_{0,u}\right)'$. 
In the sequel, for all $u\in\{1,\ldots,U\}$, we assume that $R_{0,u}>1$ and $0<p_u<1$ are random parameters. We also set $s_{0,u} + i_{0,u} = 1$ (which means that the initial number of recovered individuals is zero), with $0<i_{0,u}<1$ being a random parameter. Moreover, we consider that the infectious period $d_u = d>0$ is a fixed parameter since the duration of the infectious period can reasonably be assumed constant between different epidemics. It is important to note that the case study is outside the scope of the exponential model since a fixed parameter has been included. We refer the reader to Appendix C for implementation details.\\

\noindent Four fixed effects $\beta \in {\mathbb R}^4$ and three random effects $\xi_u=(\xi_{1,u},\xi_{3,u},\xi_{4,u})' \sim {\mathcal N}_3 (0,\Gamma)$ are considered. Therefore, using (\ref{parms_indiv}) and (\ref{transfo}), we assume the following model for the fixed and random parameters: 

\begin{equation}\label{phiSIR}
\phi_u=\left(R_{0,u},d_u,p_u,i_{0,u}\right)' = h(\beta,\xi_u), \quad  \mbox{ with }
\end{equation}
\begin{align*}
 h_1(\beta,\xi_u) &= \exp\left[\beta_1+\xi_{1,u}\right]+1,\\
 h_2(\beta,\xi_u) &= \exp\left[\beta_2\right], \\
h_i(\beta,\xi_u) &= \frac{1}{1+\exp\left[-(\beta_i+\xi_{i,u})\right]}, \ i=3,4.
\end{align*}
In other words, random effects on $(R_0,p,i_0)$ and fixed effect on $d$ are considered. Moreover, these random effects come from a priori independent sources, so that there is no reason to consider correlations between $\xi_{1,u}$, $\xi_{3,u}$ and $\xi_{4,u}$, and we can assume in this set-up a diagonal form for the  covariance matrix $\Gamma= \mbox{diag }\Gamma_{i}$, $i\in \{1,3,4\}$. 

\paragraph{Parameter values}

We consider two settings (denoted respectively (i) and (ii) below) corresponding to two levels of inter-epidemic variability (resp. high and moderate). The fixed effects values $\beta$ are chosen such that the intrinsic stochasticity of the epidemic dynamics is significant (a second set of fixed effects values leading to a lower intrinsic stochasticity is also considered; see Appendix D for details).

\begin{itemize}
  \item Setting (i): $\beta=(-0.81,0.92,1.45,-2.20)'$ and $\Gamma=\text{diag}(0.47^2,1.50^2,0.75^2)$ corresponding to $\mathbb{E}\left(R_{0,u}\right)= 1.5$, $CV_{R_{0,u}} = 17  \%$; $d=2.5$; $\mathbb{E}\left(p_{u}\right) \approx 0.74$, $CV_{p_u} \approx 31  \%$; $\mathbb{E}\left(i_{0,u}\right) \approx 0.12$, $CV_{i_{0,u}} \approx 66  \%$; \\ 
   \item Setting (ii): $\beta=(-0.72,0.92,1.45,-2.20)'$ and $\Gamma=\text{diag}(0.25^2,0.90^2,0.50^2)$ corresponding to $\mathbb{E}\left(R_{0,u}\right) = 1.5$, $CV_{R_{0,u}} = 8  \%$; $d=2.5$; $\mathbb{E}\left(p_{u}\right)\approx 0.78$, $CV_{p_u} \approx 18  \%$; $\mathbb{E}\left(i_{0,u}\right) \approx 0.11$, $CV_{i_0} \approx 45  \%$; \\
\end{itemize}
where $CV_{\phi}$ stands for the coefficient of variation of a random variable $\phi$. Let us note that the link between $\phi_u$ and $(\beta,\xi_u)$ for $p$ and $i_0$ does not have an explicit expression.

\paragraph{Data simulation} The population size is fixed to $N_u=N=10,000$. For each $U\in \{20,50,100\}$,
$J=100$ data sets, each composed of $U$ SIR epidemic trajectories,
are simulated.
Independent samplings of $\left(\phi_{u,j}=\left(R_{0,u},d_u,p_u,i_{0,u}\right)_j'\right)$, $u=1,\dots U$, $j=1,\ldots,J$, are first drawn according to model \eqref{phiSIR}. Then, conditionally to each parameter set $\phi_{u,j}$, a bidimensionnal Markov jump process $\mathcal{Z}_{u,j}(t) = (S_{u,j}(t),I_{u,j}(t))'$ is simulated. Normalizing $\mathcal{Z}_{u,j}(t)$ with respect to $N_u$ and extracting the values of the normalized process at regular time points $t_k = k \Delta$, $k=1,\ldots,n_{u,j}$, gives the $X_{u,k,j}=\left(\frac{S_{u,k,j}}{N_u},\frac{I_{u,k,j}}{N_u}\right)'$'s. A fixed discretization time step is used, {\it i.e.} the same value of $\Delta$ is used to simulate all the epidemic data. For each epidemic, $T_{u,j}$ is defined as the first time point at which the number of infected individuals becomes zero. Two values of $\Delta$ are considered ($\Delta\in\{0.425,2\}$) corresponding to an average number of time-point observations $\overline{n}_j = \frac{1}{U}\sum_{u=1}^{U} n_{u,j} \in\{20,100\}$.
Only trajectories that did not exhibit early extinction were considered for inference. The theoretical proportion of these trajectories is given by $1-(1 / R_0)^{I_0}$ (\cite{Andersson2000}). Then, given the simulated $X_{u,k,j}$'s and parameters $\phi_{u,j}$'s, the observations $Y_{u,k,j}$ are generated from binomial distributions $\mathcal{B}(I_{u,k,j},p_{u,j})$.



\subsection{Point estimates and standard deviations for inferred parameters}

Tables \ref{exp1_CV2} and \ref{exp1_CV1} show the estimates of the expectation and standard deviation of the mixed effects $\phi_{u}$, computed from the estimations of $\beta$ and $\Gamma$ using functions $h$ defined in (\ref{phiSIR}), for settings (i) and (ii). For each parameter, the reported values are the mean of the $J=100$ parameter estimates $\phi_{u,j}$, $j\in \{1,\ldots,J\}$, and their standard deviations in brackets. \\

{\setlength{\tabcolsep}{6pt}
\begin{table}[h]
\begin{center}
\caption{Estimates for setting (i): high inter-epidemic variability. For each combination of $(\overline{n},U)$ and for each model parameter (defined in the first line of the table), point estimates and precision are calculated as the mean of the $J=100$ individual estimates and their standard deviations (in brackets).} \label{exp1_CV2}
\small
\begin{tabular}{cc|ccccccc}
Parameters & &  $\mathbb{E}\left(R_{0,u}\right)$ & $d$  & $\mathbb{E}\left(p_{u}\right)$  & $\mathbb{E}\left(i_{0,u}\right)$ & 
$\text{sd}\left(R_{0,u}\right)$ & $\text{sd}\left(p_{u}\right)$ & $\text{sd}\left(i_{0,u}\right)$ \\
\hline
True values & & \textbf{1.500} & \textbf{2.500} & \textbf{0.739} & \textbf{0.119} & \textbf{0.250} & \textbf{0.226} & \textbf{0.079} \\
 \hline
  & & & & & & & &  \\
 $\overline{n}=20$ & $U=20$  & 1.580  &  2.584  &  0.688  &  0.126  &  0.335  &  0.193  &  0.078  \\
  & &  (\textit{0.135}) & (\textit{0.293}) & (\textit{0.117}) & (\textit{0.024}) & (\textit{0.151}) & (\textit{0.051}) & (\textit{0.020}) \\
    & & & & & & & &  \\
 & $U=50$  &  1.574  &  2.538  &  0.704  &  0.122  &  0.359  &  0.201  &  0.079  \\
 & &  (\textit{0.111}) & (\textit{0.220}) & (\textit{0.089}) & (\textit{0.019}) & (\textit{0.149}) & (\textit{0.030}) & (\textit{0.014}) \\
   & & & & & & & &  \\
  & $U=100$   & 1.583  &  2.564  &  0.700  &  0.124  &  0.385  &  0.199  &  0.081  \\
 &   &  (\textit{0.105}) & (\textit{0.210}) & (\textit{0.083}) & (\textit{0.015}) & (\textit{0.134}) & (\textit{0.023}) & (\textit{0.011}) \\
   & & & & & & & &  \\
 \hline
  & & & & & & & &  \\
$\overline{n}=100$ & $U=20$   & 1.501  &  2.502  &  0.734  &  0.118  &  0.292  &  0.217  &  0.075  \\
  & & (\textit{0.080}) & (\textit{0.159}) & (\textit{0.059}) & (\textit{0.021}) & (\textit{0.105}) & (\textit{0.035}) & (\textit{0.019}) \\
    & & & & & & & &  \\
 & $U=50$  &  1.510  &  2.522  &  0.729  &  0.120  &  0.305  &  0.217  &  0.080  \\
 & &  (\textit{0.054}) & (\textit{0.126}) & (\textit{0.038}) & (\textit{0.014}) & (\textit{0.070}) & (\textit{0.022}) & (\textit{0.012}) \\
   & & & & & & & &  \\
  & $U=100$   & 1.503  &  2.508  &  0.738  &  0.119  &  0.308  &  0.216  &  0.079  \\
 &   &  (\textit{0.047}) & (\textit{0.097}) & (\textit{0.030}) & (\textit{0.010}) & (\textit{0.054}) & (\textit{0.016}) & (\textit{0.009}) \\
\end{tabular}
\end{center}
\end{table}
}

{\setlength{\tabcolsep}{6pt}
\begin{table}[h]
\begin{center}
\caption{Estimates for setting (ii): moderate inter-epidemic variability. For each combination of $(\overline{n},U)$ and for each model parameter (defined in the first line of the table), point estimates and precision are calculated as the mean of the $J=100$ individual estimates and their standard deviations (in brackets).} \label{exp1_CV1}
\small
\begin{tabular}{cc|ccccccc}
Parameters & &  $\mathbb{E}\left(R_{0,u}\right)$ & $d$  & $\mathbb{E}\left(p_{u}\right)$  & $\mathbb{E}\left(i_{0,u}\right)$ & 
$\text{sd}\left(R_{0,u}\right)$ & $\text{sd}\left(p_{u}\right)$ & $\text{sd}\left(i_{0,u}\right)$ \\
\hline
True values & & \textbf{1.500} & \textbf{2.500} & \textbf{0.777} & \textbf{0.109} & \textbf{0.125} & \textbf{0.143} & \textbf{0.049} \\
 \hline
  & & & & & & & &  \\
 $\overline{n}=20$ & $U=20$  & 1.619  &  2.764  &  0.666  &  0.127  &  0.190  &  0.117  &  0.053  \\
  & &  (\textit{0.120}) & (\textit{0.256}) & (\textit{0.099}) & (\textit{0.022}) & (\textit{0.106}) & (\textit{0.034}) & (\textit{0.014}) \\
    & & & & & & & &  \\
 & $U=50$  &  1.638  &  2.789  &  0.653  &  0.128  &  0.213  &  0.122  &  0.056  \\
 & &  (\textit{0.103}) & (\textit{0.233}) & (\textit{0.087}) & (\textit{0.018}) & (\textit{0.099}) & (\textit{0.018}) & (\textit{0.010}) \\
   & & & & & & & &  \\
  & $U=100$   & 1.623  &  2.769  &  0.658  &  0.128  &  0.209  &  0.122  &  0.056  \\
 &   &  (\textit{0.081}) & (\textit{0.194}) & (\textit{0.075}) & (\textit{0.013}) & (\textit{0.056}) & (\textit{0.017}) & (\textit{0.007}) \\
   & & & & & & & &  \\
 \hline
  & & & & & & & &  \\
$\overline{n}=100$ & $U=20$   & 1.540  &  2.627  &  0.732  &  0.118  &  0.176  &  0.143  &  0.050  \\
  & &  (\textit{0.066}) & (\textit{0.143}) & (\textit{0.057}) & (\textit{0.017}) & (\textit{0.055}) & (\textit{0.035}) & (\textit{0.012}) \\
    & & & & & & & &  \\
 & $U=50$  & 1.539  &  2.622  &  0.733  &  0.117  &  0.183  &  0.145  &  0.052  \\
 & &  (\textit{0.044}) & (\textit{0.098}) & (\textit{0.041}) & (\textit{0.009}) & (\textit{0.038}) & (\textit{0.018}) & (\textit{0.007}) \\
   & & & & & & & &  \\
  & $U=100$   & 1.541  &  2.629  &  0.732  &  0.118  &  0.187  &  0.149  &  0.053  \\
 &   &  (\textit{0.040}) & (\textit{0.078}) & (\textit{0.030}) & (\textit{0.008}) & (\textit{0.035}) & (\textit{0.016}) & (\textit{0.006}) \\
\end{tabular}
\end{center}
\end{table}
}

\noindent The results show that all the point estimates are close to the true values (relatively small bias), whatever the inter-epidemic variability setting, even for small values of $\bar{n}$ and $U$.
When the number of epidemics $U$ increases, the standard error of the estimates decreases, but it does not seem to have a real impact on the estimation bias. Besides, observations of higher frequency of the epidemics (large $\bar{n}$) lead to lower bias and standard deviations. It is particularly marked concerning both expectation and standard deviations of the random parameters $R_{0,u}$ and $p_{u}$. Irrespective to the level of inter-epidemic variability, the estimations are quite satisfactory. While standard deviations of $R_{0,u}$ are slightly over-estimated, even for large $U$ and $\overline{n}$, this trend in bias does not affect the standard deviations of $p_{u}$ and $i_{0,u}$. \\

\begin{figure}[H]
	\includegraphics[width=0.95\textwidth,height=10.5cm]{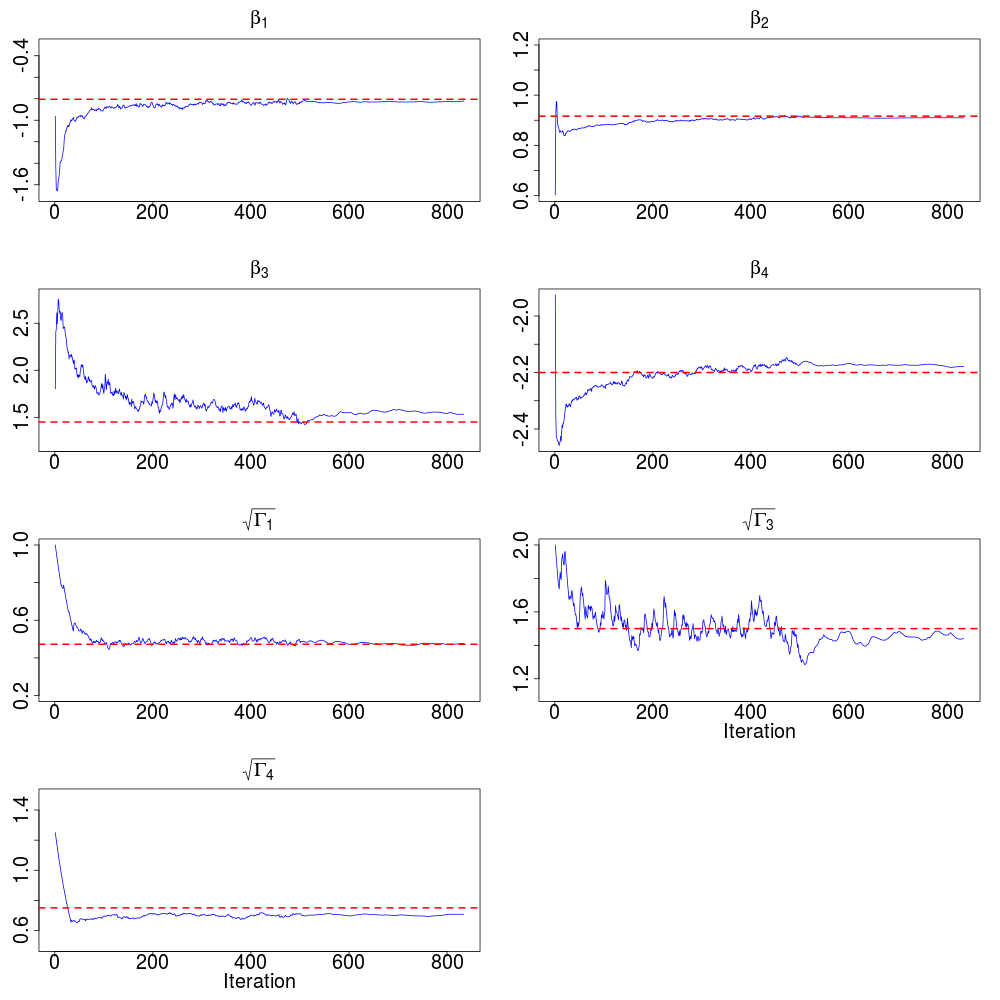}\\
	\caption{Convergence graphs of the SAEM algorithm for estimates of $\beta=(\beta_1,\beta_2,\beta_3,\beta_4)$ and $\text{diag}(\Gamma)=(\Gamma_1,\Gamma_3,\Gamma_4)$. Setting (i) with $U=100$ and $\bar{n}=100$. Parameter values at each iteration of the SAEM algorithm (plain blue line) and true values of model parameters (dotted red line).}\label{CV_exp1}
\end{figure}

\noindent For a given data set, Figure \ref{CV_exp1} displays convergence graphs of the SAEM algorithm for each estimates of model parameters in setting (i) with $U=100$ and $\bar{n}=100$. Although the model does not belong to the curved exponential family, convergence of model parameters towards their true value is obtained for all parameters. 

\subsection{Comparison with an empirical two-step approach}

The inference proposed method (referred to as SAEM-KM) is compared to an empirical two-step approach not taking into account explicitly mixed effects in the model. For that purpose, let us consider the method presented in (\cite{Narci2020}) (referred to as KM) performed in two steps: first, we compute the estimates $\hat{\phi}_u$ independently on each of the $U$ trajectories. Second, the empirical mean and variance of the $\hat{\phi}_{u}$'s are computed. We refer the reader to Appendix C for practical considerations on implementation of the KM method.  \\



\begin{figure}[h]
	\includegraphics[width=0.95\textwidth,height=12.5cm]{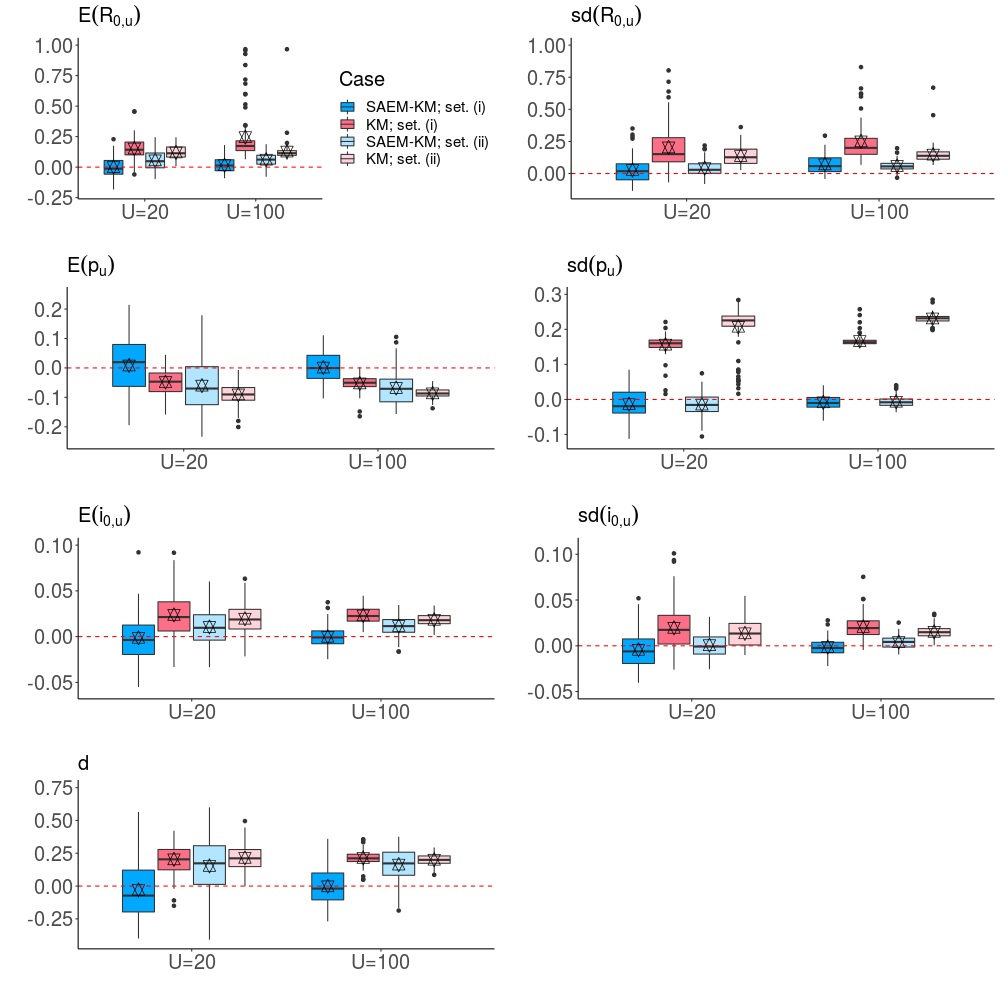}\\
	\caption{Boxplots (25th, 50th and 75th percentiles) of the bias of the estimates of each model parameter, with $\bar{n}=50$, obtained with SAEM-KM (blue boxes) and KM (red boxes). Two levels: $U=20$ and $U=100$ epidemics. Dark colours: high inter-epidemic variability (setting (i)). Light colours: moderate inter-epidemic variability (setting (ii)). The symbol represents the estimated mean bias. For sake of clarity, we removed extreme values from the graphical representation. This concerns only the parameter $R_0$ and the KM method: $37$ values for $\mathbb{E}(R_{0,u})$ ($35$ in setting (i), $2$ in setting (ii)) and $50$ values for $\text{sd}(R_{0,u})$ ($47$ in setting (i), $3$ in setting (ii)).} \label{boxplot_exp1}
\end{figure}

\noindent Let us consider $\bar{n}=50$ and $U\in \{20,100\}$. Figure \ref{boxplot_exp1} displays the distribution of the bias of the parameter estimates $\phi_{u,j}$, $j\in \{1,\ldots,J\}$, $J=100$, obtained with SAEM-KM and KM for simulation settings (i) and (ii).

\noindent We notice a clear advantage to consider the mixed-effects structure. Overall, the results show that SAEM-KM outperforms KM. This is more pronounced for standard deviation estimates in the large inter-epidemic variability setting (i) than in the moderate inter-epidemic variability setting (ii). Concerning the expectation estimates, their dispersion around the median is lower for KM than for SAEM-KM, especially in setting (ii), but the bias of KM estimates is also higher. When the inter-epidemic variability is high (setting (i)), the performances of the two inference methods are substantially different. In particular, KM sometimes fails to provide plausible estimates (especially for parameter $R_0$).\\

\noindent We also tested other values for $\bar{n}$ and $N$ (not shown here), e.g. $\bar{n}=20$ (lower amount of information) and $N=2000$ (higher intrinsic variability of epidemics). In such cases, KM also failed to provide satisfying estimations whereas the mixed-effects approach was much more robust.   

\section{Case study: influenza outbreaks in France}
\label{sec:estimrealdata}

\paragraph{Data} The SAEM-KM method is evaluated on a real data set of influenza outbreaks in France provided by the Réseau Sentinelles (url: \url{www.sentiweb.fr}). We use the daily number of influenza-like illness (ILI) cases between 1990-2017, considered as a good proxy of the number of new infectious individuals. The daily incidence rate was expressed per $100,000$ inhabitants. To select epidemic periods, we chose the arbitrary threshold of weekly incidence of $160$ cases per $100,000$ inhabitants (\cite{Cauchemez2008}), leading to $28$ epidemic dynamics. Two epidemics have been discarded due to their bimodality (1991-1992 and 1998). Therefore, $U=26$ epidemic dynamics are considered for inference.  \\ 

\paragraph{Compartmental model} 

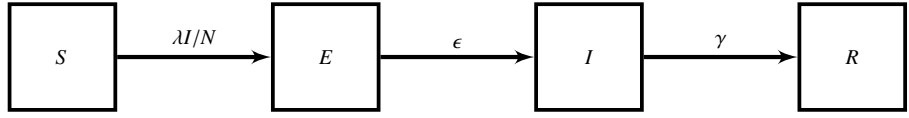
\begin{figure}[H] 
\centering
\begin{tikzpicture}[node distance=3cm,auto,>=latex']
    \node[int] (c) [] {$S$};

    \node [int] (d) [right of=c, node distance=3.5cm] {$E$};

    \node [int] (e) [right of=d, node distance=3.5cm] {$I$};
    
    \node [int] (f) [right of=e, node distance=3.5cm] {$R$};

    \draw[->, ultra thick, black] (c) edge node {$\lambda I/N$} (d);
    \draw[->, ultra thick, black] (d) edge node {$\epsilon$} (e);
    \draw[->, ultra thick, black] (e) edge node {$\gamma$} (f);
\end{tikzpicture}
\caption{SEIR compartmental model with four blocks corresponding respectively to susceptible (S), exposed (E), infectious (I) and recovered (R) individuals. Transitions of individuals
from one health state to another are governed by the transmission rate $\lambda$, the incubation rate $\epsilon$ and the recovery rate $\gamma$.}
\label{fig:SEIR_rates}
\end{figure}

Let us consider the SEIR model (see Figure \ref{fig:SEIR_rates}). An individual is considered exposed (E) when infected but not infectious. Denote $\eta=(\lambda,\epsilon,\gamma,x_0)$, with $x_0=(s_0,e_0,i_0,r_0)$, the parameters involved in the transition rates, where $\epsilon$ is the transition rate from $E$ to $I$. ODEs of the SEIR model are as follows:

\begin{align}
\label{SEIRODE}
 \begin{cases}
  \frac{ds}{dt}(\eta,t) &= - \lambda s(\eta,t) i(\eta,t), \\
  \frac{de}{dt}(\eta,t) &= \lambda s(\eta,t) i(\eta,t) - \epsilon e(\eta,t), \\
  \frac{di}{dt}(\eta,t) &= \epsilon e(\eta,t) -\gamma i(\eta,t), \\
  \frac{dr}{dt}(\eta,t) &= \gamma i(\eta,t), \\
  x_0 &= (s_0,e_0,i_0,r_0).
 \end{cases} 
\end{align}

\noindent Another parametrization exhibits the basic reproduction number $R_0=\frac{\lambda}{\gamma}$, the incubation period $d_E=\frac{1}{\epsilon}$ and the infectious period $d_I=\frac{1}{\gamma}$. Thus, the epidemic pamareters are $\eta=(R_0,d_E,d_I,s_0,e_0,i_0)'$. Let us describe the two-layer model used in the sequel. 

\paragraph{Intra-epidemic variability} For each epidemic $u$, let $X_u=\left(\frac{S_u}{N_u},\frac{E_u}{N_u},\frac{I_u}{N_u}\right)'$ and $$\eta_u = \left(R_{0,u},d_{E,u},d_{I,u},s_{u,0},e_{u,0},i_{u,0}\right),$$ where the population size is fixed at $N_u = N = 100,000$. Denote by $\text{Inc}_u(t_k)$ the number of newly infected individuals at time $t_k$ for epidemic $u$. We have
\begin{equation*}\label{NI}
\text{Inc}_u(t_k)= \int_{t_{k-1}}^{t_k} \frac{1}{d_{E,u}} E_u(t) dt=S_u(t_{k-1}) - S_u(t_k) + E_u(t_{k-1}) - E_u(t_k)= - (\Delta_k S_u+\Delta_k E_u).\end{equation*}
Observations are modeled as incidence data observed with Gaussian noises. We draw our inspiration from (\cite{Breto2018}) to account for over-dispersion in data. Therefore, assuming a reporting rate $p_u$ for epidemic $u$, the mean and the variance of the observed newly infected individuals are respectively defined as $p_u \text{Inc}_u(t_k)$ and $p_u \text{Inc}_u(t_k) + \tau_u^2 p_u^2 \text{Inc}_u(t_k)^2$, where parameter $\tau_u$ is introduced to handle over-dispersion in the data. Denote $\phi_u = \left(\eta_u,p_u,\tau_u^2\right)$. Therefore, we use the model defined in (\ref{Kalman_gen_inc})
with $\Delta_k X_u = \left(\frac{\Delta_k S_u}{N},\frac{\Delta_k E_u}{N},\frac{\Delta_k I_u}{N}\right)'$, $V_{u,k} \sim  \mathcal{N}_d\left(0, T_k(\phi_u,\Delta)\right)$,  $\tilde{W}_{u,k} \sim {\cal N} _q( 0, \tilde{P}_k(\phi_u))$, $G_k(\cdot)$, $A_{k-1}(\cdot)$ and $T_k(\cdot)$ deriving from \eqref{derive_SEIR} in Appendix \ref{quantities_seir}, $\tilde{B}(\phi_u) = (-p_u \ \ -p_u \ \ 0)$ and 
$$\tilde{P}_k(\phi_u) = \frac{1}{N} \left(\tilde{B}(\phi_u) \Delta_k x_u +   \tau_u^2 \left(\tilde{B}(\phi_u) \Delta_k x_u\right) ^2 \right),$$ where $x(\cdot,t)$ is the ODE solution of (\ref{SEIRODE}). 


\paragraph{Inter-epidemic variability} Let us first comment on the duration of the incubation period $d_E$ and of the infectious period $d_I$. Studies in the literature found discrepant values of these durations (see \cite{Cori2012} for a review), varying from $0.64$ (\cite{Fraser2009}) to $3.0$ (\cite{Pourbohloul2009}) days for the incubation period and from $1.27$ (\cite{Fraser2009}) to $8.0$ (\cite{Pourbohloul2009}) days for the infectious period. For example, \cite{Cori2012} estimated that $d_E=1.6$ and $d_I=1.0$ days on average using excretion profiles from experimental infections. In two other papers, these durations were fixed according to previous studies (e.g. \cite{Mills2004}, \cite{Ferguson2005}): $(d_E,d_I) = (1.9,4.1)$ days (\cite{Chowell2008}); $(d_E,d_I) = (0.8,1.8)$ days (\cite{Baguelin2013}). Performing a systematic review procedure from viral shedding and/or symptoms, \cite{Carrat2008} estimated $d_E$ to be between $1.7$ and $2.0$ on average. For identifiability reasons, we consider the latent and infectious periods $d_E$ and $d_I$ known and test three combinations of values: $(d_E,d_I) = (1.6,1.0)$, $(0.8,1.8)$ and $(1.9,4.1)$. \\

\noindent We consider that the basic reproduction number $R_0$ and the reporting rate $p$ are random, reflecting the assumptions that the transmission rate of the pathogen varies from season to season and the reporting could change over the years. Moreover, we assume $e_u(0)=i_u(0)$ random and unknown (\textit{i.e.} the proportion of initial exposed and infectious individuals is variable between epidemics). \cite{Cauchemez2008} assumed that at the start of each influenza season, a fixed average of $27\%$ of the population is immune, that is $r_{0,u}=r_0=0.27$. To assess the robustness of the model with respect to the $r_0$ value, we test three values: $r_0 \in \{0.1,0.27,0.5\}$. This leads to $s_{0,u} = 1 - r_0 - 2 i_{0,u}$ random and unknown. Finally, we assume that $\tau_u^2=\tau^2$ is fixed and unknown. To sum up, we have to consider in the model: known parameters $(d_E,d_I)\in \{(0.8,1.8),(1.6,1.0),(1.9,4.1)\}$ and $r_0\in \{0.1,0.27,0.5\}$; fixed and unknown parameter $\tau^2$ ; random and unknown parameters $R_0$, $i_{0}$ and $p$. \\


\noindent Therefore, using \eqref{parms_indiv}, we consider the following model for random parameters:
\begin{equation*}
\phi_u=\left(R_{0,u},p_u,i_{0,u},\tau^2\right)' = h(\beta,\xi_u), \quad  \mbox{ with }
\end{equation*}
\begin{align*}
  h_1(\beta,\xi_u) &= \exp\left[\beta_1+\xi_{1,u}\right]+1,\\
h_j(\beta,\xi_u) &= \frac{1}{1+\exp\left[-(\beta_j+\xi_{j,u})\right]}, \ j=2,3, \\
  h_4(\beta,\xi_u) &= \exp\left[\beta_4\right], 
\end{align*} 
where fixed effects $\beta \in \mathbb{R}^4$ and the random effects are  $\xi_u \sim_{i.i.d.} \mathcal{N}_3(0,\Gamma)$ with $\Gamma$ a covariance matrix assumed to be diagonal.
 
\paragraph{Parameter estimates}  We consider nine models with different combinations of values of $((d_E,d_I),r_0)$. Using importance sampling techniques, we estimate the observed log-likelihood of each model from the estimated parameters values initially obtained with the SAEM algorithm. Table \ref{log_lik} provides the estimated log-likelihood values of the nine models of interest. Irrespectively of the $r_0$ value, we find that the model with $(d_E,d_I)=(1.9,4.1)$ outperforms the two other models in terms of log-likelihood value. Moreover, for a given combination of values of $(d_E,d_I)$, the estimated log-likelihood values are quite similar according to the three $r_0$ tested values.

\begin{table}[H]
\caption{Estimated values of the observed log-likelihood of the model obtained by testing nine combinations of values of $((d_E,d_I),r_0)$.}
\label{log_lik}
    \centering
    \begin{tabular}{c|cc}
    $(d_E,d_I)$ & $r_0$ & Estimated log-likelihood \\
    \hline
    (0.8,1.8) & 0.1 & 9011.752 \\ 
              & 0.27 & 8827.870 \\
              & 0.5 & 8499.452 \\
              & & \\
    (1.6,1.0) & 0.1 & 9147.108 \\
              & 0.27 & 8961.991 \\
              & 0.5 & 8643.562 \\
              & & \\
    (1.9,4.1) & 0.1 & 10270.000 \\
              & 0.27 & 10216.260 \\
              & 0.5 & 9905.436 \\
    \end{tabular}
\end{table}

\noindent Let us focus on the model with $(d_E,d_I)=(1.9,4.1)$. Table \ref{res_real} presents the estimation results of the model parameters obtained by testing the three values of $r_0$: $0.1$, $0.27$ and $0.5$.

\begin{table}[H]
\caption{Estimates of the mean, 5th and 95th percentiles and coefficient of variation (CV) for model parameters $\left(R_{0,u},i_{0,u},p_u,\tau^2\right)'$, assuming $(d_E,d_I)=(1.9,4.1)$ and testing three values of $r_0$: $0.1$, $0.27$ and $0.5$. For fixed parameter, only the estimated mean is available.}
\label{res_real}
    \centering
    \begin{tabular}{cc|cccc}
   & & $R_{0,u}$ & $p_{u}$ & $i_{0,u}$  & $\tau^2$ \\
    \hline
    Estimated mean & $r_0=0.1$ & 1.810 & 0.069 & 0.010 & 0.025 \\
                               & $r_0=0.27$ & 2.238 & 0.084 & 0.008 & 0.013  \\
                               & $r_0=0.5$ & 3.281 & 0.119 & 0.006 & 0.037  \\
                                   & & & & \\
    Estimated [5th,95th] percentiles & $r_0=0.1$ & [1.470,2.264] & [0.026,0.138] & [0.003,0.023] & --- \\
                               & $r_0=0.27$ & [1.787,2.825] & [0.031,0.169] & [0.002,0.019] & --- \\
                               & $r_0=0.5$ & [2.696,3.977] & [0.044,0.238] & [0.002,0.014] & --- \\
                                   & & & & \\
    Estimated $CV$ & $r_0=0.1$ &  14 \% & 53 \% & 67 \% & --- \\
                               & $r_0=0.27$ & 14 \% & 52 \% & 72 \% & --- \\
                               & $r_0=0.5$ & 12 \% & 51 \% & 74 \% & --- \\
    \end{tabular}
\end{table}

\noindent The average estimated value of $R_0$ is quite contrasted according to the $r_0$ value: between $1.81$ and $3.28$ from $r_0=0.1$ to $r_0=0.5$. By comparison, in (\cite{Cauchemez2008}), $R_0$ is estimated to be $1.7$ during school term, and $1.4$ in holidays, using a population structured into households and schools. \cite{Chowell2008} estimated a different reproduction number $\tilde{R}=(1-r_0)R_0=1.3$, measuring the transmissibility at the beginning of an epidemic in a partially immune population, from mortality data. In our case, the average value of $\tilde{R}$ is estimated to $1.63$, $1.63$ and $1.64$ when $r_0=0.1$, $0.27$ and $0.5$ respectively. Therefore, given the nature of the observations (new infected individuals) and the considered model, this appears to be difficult to correctly identify $R_0$ together with $r_0$. Indeed, the fraction of immunized individuals at the beginning of each seasonal influenza epidemic is an important parameter for the epidemic dynamics, but its value is not well known. This has implications for the stability of the estimation of the other parameters.
Interestingly, the average reporting rate is estimated particularly low (around $10 \%$ irrespective of the $r_0$ value). Moreover, we observe that $R_0$ together with $p$ and $i_0$ seem to be variable from season to season, with moderate coefficient of variation $\text{CV}(R_{0,u})$ close to $15\%$ and high coefficients of variation $\text{CV}(p_{u})$ and $\text{CV}(i_{0,u})$ around $50 \%$ and $70 \%$ respectively. 

   \begin{figure}[H]
	\includegraphics[width=0.95\textwidth,height=7.5cm]{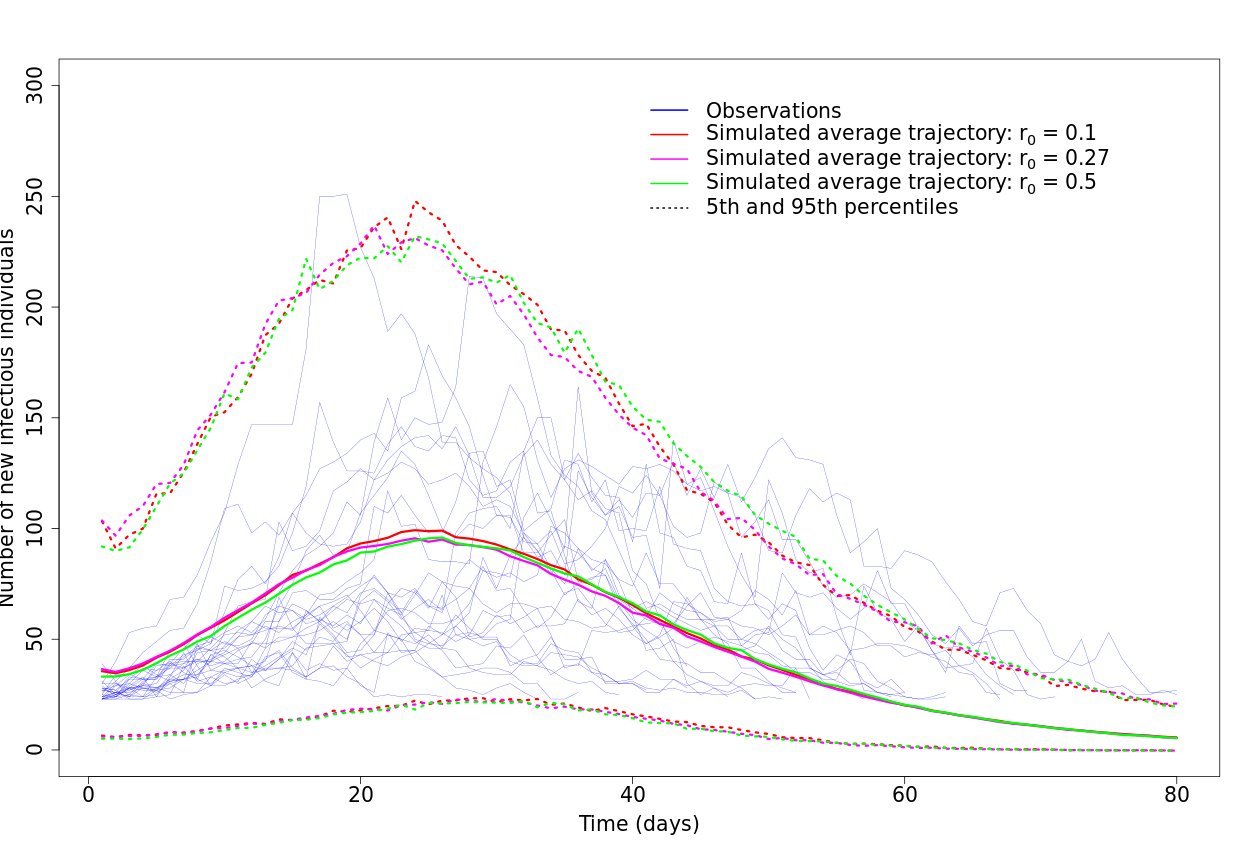}\\
	\caption{Post-predictive check. Observations (number of ILI as proxy for new infectious for each of the $U$ epidemics) (blue). Simulated trajectories obtained for $r_0=0.1$ (red), $r_0=0.27$ (magenta) and $r_0=0.5$ (green) in three steps: (i) generation of $1000$ $\hat{\phi}_u$ values based on estimated values of parameters; (ii) given $\hat{\phi}_u$, simulation of $1000$ epidemics according to the model (\ref{Kalman_gen_inc}); (iii) computation of average trajectory (solid line) and 5th and 95th percentiles (dotted lines) of the $1000$ simulated epidemics. Population size fixed to $N=100,000$.}\label{SAEM_inc160}
\end{figure}

\noindent The post-predictive check is shown in Figure \ref{SAEM_inc160}. The difference between the average simulated curves obtained with estimated parameter values is negligible according to the $r_0$ value. Considering the values of $\tilde{R}$, very close in the three scenarios, the proximity of the predicted trajectories is not surprising. Let us emphasize that the majority of the observations are within the predicted envelope (5th and 95th percentiles). Moreover, the predicted average trajectory informs about generic trends of influenza outbreaks: on average, the epidemic peak should be reached around $25$ days after the beginning of the outbreak with an incidence of $90/100,000$ inhabitants approximately.

\section{Discussion}
\label{sec:discussion}

In this paper, we propose a generic inference method taking into account simultaneously in a unique model multiple epidemic trajectories and providing estimations of key parameters from incomplete and noisy epidemic data (prevalence or incidence). The framework of the mixed-effects models was used to describe the inter-epidemic variability, whereas the intra-epidemic variability was modeled by an autoregressive Gaussian process. The Gaussian formulation of the epidemic model for prevalence data used in (\cite{Narci2020}) was extended to the case where incidence data were considered. Then, the SAEM algorithm was coupled with Kalman-like filtering techniques in order to estimate model parameters. \\

\noindent
The performances of the estimators were investigated on simulated data of SIR dynamics, under various scenarios, with respect to the parameter values of epidemic and observation processes, the number of epidemics ($U$), the average number of observations for each of the $U$ epidemics ($\bar{n}$) and the population size ($N$).  The results show that all estimates are close to the true values (reasonable biases), whatever the inter-epidemic variability setting, even for small values of $\bar{n}$ and $U$. The performances, in term of precision, are improved when increasing $U$, whereas the bias and standard deviations of the estimations decrease when increasing $\bar{n}$. We also compared our method with a two-step empirical approach that processes the different data sets separately and combines the individual parameter estimates a posteriori to provide an estimate of inter-epidemic variability (\cite{Narci2020}). When the number of observations is too low and/or the coefficient of variation of the random effects is high, SAEM-KM clearly outperforms KM. \\

\noindent The proposed inference method was also evaluated on an influenza data set provided by the Réseau Sentinelles, consisting in the daily number of new infectious individuals per $100,000$ inhabitants between 1990 and 2017 in France, using a SEIR compartmental model. Testing different combinations of values for $(d_E,d_I)$ and $r_0$, we find that $(d_E,d_I)=(1.9,4.1)$ leads to the best fitting model.
Then, irrespective to the $r_0$ value, we estimated an average value of $\tilde{R}=(1-r_0)R_0$ to be around $1.6$. Moreover, we highlighted a non-negligible variability from season to season that is quantitatively assessed. This variability appears especially in the initial conditions ($i_0$) and the reporting rate ($p$), as a combined effect of observational uncertainties and differences between seasons. Although to a lesser extent, $R_0$ also appears to vary between seasons, plausibly reflecting the variability in the transmission rate ($\lambda$). Obviously, the estimations can strongly depend on the choice of the compartmental model, the nature and frequency of the observations and the distribution of the random parameters. Our contribution is to propose a finer estimation of the model parameters by taking into account simultaneously all the influenza outbreaks in France for the inference procedure. This leads to an explicit and rigorous estimation of the seasonal variability. \\



\noindent Other methods have been implemented to deal with multiple epidemic dynamics. \cite{Breto2020} proposed a likelihood-based inference methods for panel data modeled by non-linear partially observed jump processes incorporating unit-specific parameters and shared parameters. Nevertheless, the framework of mixed-effects models was not really investigated. \cite{prague:hal-02555100} used an ODE system with mixed effects on the parameters to analyse the first epidemic wave of Covid-19 in various regions in France by inferring key parameters from the daily incidence of infectious ascertained and hospitalized infectious cases. 
To our knowledge, there are no published studies aiming at the estimation of key parameters simultaneously from several outbreak time series using both a stochastic modeling of epidemic processes and random effects on model parameters. \\

\noindent The main advantage of our method is to propose a direct access to the inter-epidemic variability between multiple outbreaks. Taking into account simultaneously several epidemics in a unique model leads to an improvement of statistical inference compared with empirical methods which consider independently epidemic trajectories. For example, we can mention two experimental settings: (1) the number of epidemics is high but the number of observations per epidemic is low; (2) the number of observations per epidemic is high but the number of epidemics is low. In such cases, mixed-effects approaches can provide more satisfying estimation results. This benefit more than compensates for the careful calibration of the tuning parameters of the SAEM algorithm. \\

\noindent In some practical cases in epidemiology, it might be difficult to determine whether a parameter is fixed or random. Consequently, our approach could be associated with model selection techniques to inform this choice, using a criterion based on the log-likelihood of observations (see for instance (\cite{Delattre2014}) and (\cite{Delattre2020})). This would allow to determine more precisely which parameters reflect inter-individual variability and thus help to better understand the mechanisms underlying this variability.  Moreover, we presented a case study on influenza outbreaks, where the variability between epidemics is seasonal, but our approach can be also applied on epidemics spreading simultaneously in many regions. In this case, the inter-epidemic variability is spatial and it would be interesting to evaluate trends from one region to another.

\section*{Funding}

This work was supported by the French Agence National de la Recherche [project CADENCE, ANR-16-CE32-0007-01] and by a grant from Région Île-de-France (DIM MathInnov). 

\section*{Acknowledgments} We thank the Réseau Sentinelles (INSERM/Sorbonne Université, \url{www.sentiweb.fr}) for providing a real data set of influenza outbreaks in France.

\renewcommand{\thesection}{\Alph{section}}
\setcounter{section}{0}
\appendix

\normalsize

\section{Key quantities involved in the SEIR epidemic model}
\label{quantities_seir}

In the SEIR model, epidemic parameters are the transition rates $\lambda$, $\epsilon$ and $\gamma$ and the initial proportions of susceptible, exposed and infectious individuals $s_0=\frac{S(0)}{N}$, $e_0=\frac{E(0)}{N}$ and 
$i_0=\frac{I(0)}{N}$. When there is no ambiguity, we denote by $s$, $e$ and $i$ respectively the solutions $s(\eta,t)$, $e(\eta,t)$ and $i(\eta,t)$ of the system of ODEs defined in \eqref{SEIRODE}. Then, the functions
$b(\eta,\cdot)$ and $\Sigma(\eta,\cdot)$ are
\begin{equation}
\label{derive_SEIR}
 b(\eta, s,e,i) = \begin{pmatrix} - \lambda s i \\ \lambda s i -\epsilon e \\ \epsilon e - \gamma i  \end{pmatrix}; \quad  \Sigma(\eta,s,e,i) = \begin{pmatrix} \lambda s i & - \lambda s i & 0 \\ - \lambda s i & \lambda s i + \epsilon e & - \epsilon e \\ 0  & -\epsilon e & \epsilon e + \gamma i \end{pmatrix},
\end{equation}
and the Cholesky decomposition of $\Sigma (\eta,\cdot)$ yields
\begin{equation*}\label{sigeta_SEIR}
    \sigma(\eta,s,e,i) = \begin{pmatrix} \sqrt{\lambda s i} & 0 & 0 \\ - \sqrt{\lambda s i} &  \sqrt{\epsilon e} & 0 & \\ 0 & -\sqrt{\epsilon e} & \sqrt{\gamma i} \end{pmatrix}.
\end{equation*}

\section{Details on the Kalman filter equations for incidence data of epidemic dynamics} 

Consider the model (\ref{KalmanModel_inc}). Assume that 
 $\mathcal{L}(\Delta_1 X) = \mathcal{N}_d(G_1,T_1)$ and $\mathcal{L}(Y_1 | \Delta_1 X) = \mathcal{N}_q(B\Delta_1 X, P_1)$. 
Let $\widehat{\Delta_1 X} = G_1 = x(t_1) - x_0$ and $\widehat{\Xi}_1 = T_1$. Then, at iteration $k=1$, the three steps of the Kalman filter are: \\

\begin{itemize}
\item[1.] Prediction: $\mathcal{L}(\Delta_2 X|Y_1) = \mathcal{N}_d(\widehat{\Delta_2 X},\widehat{\Xi}_2)$
\begin{align*}
 \widehat{\Delta_2 X} &= G_2 + (A_1-I_d) \overline{\Delta_1 X} \\
 \widehat{\Xi}_2 &= (A_1-I_d) \overline{T}_1 (A_1-I_d)' + T_2
\end{align*}
 \item[2.] Updating: $\mathcal{L}(\Delta_1 X|Y_1) = \mathcal{N}_d(\overline{\Delta_1 X},\overline{T}_1)$
 \begin{align*}
  \overline{\Delta_1 X} &= \widehat{\Delta_1 X} + \widehat{\Xi}_1 \tilde{B}' (\tilde{B} \widehat{\Xi}_1 \tilde{B}' + \tilde{P}_1)^{-1}(Y_1 - \tilde{B}\widehat{\Delta_1 X}) \\
  \overline{T}_1 &= \widehat{\Xi}_1 - \widehat{\Xi}_1 \tilde{B}' (\tilde{B} \widehat{\Xi}_1 \tilde{B}' + \tilde{P}_1)^{-1} \tilde{B} \widehat{\Xi}_1
 \end{align*}
\item[3.] Marginal: $\mathcal{L}(Y_2|Y_1) = \mathcal{N}(\widehat{M}_2,\widehat{\Omega}_2)$
\begin{align*}
 \widehat{M}_2 &= \tilde{B} \widehat{\Delta_2 X} \\
 \widehat{\Omega}_2 &= \tilde{B} \widehat{\Xi}_2 \tilde{B}' + \tilde{P}_2
\end{align*}
\end{itemize}

\noindent Now, starting from the distribution of $\mathcal{L}(\Delta_2 X|Y_1)$, the Kalman filter at iteration $k=2$ becomes:

\begin{itemize}
\item[1.] Prediction: $\mathcal{L}(\Delta_3 X|Y_2,Y_1) = \mathcal{N}_d(\widehat{\Delta_3 X},\widehat{\Xi}_3)$
\begin{align*}
 \widehat{\Delta_3 X} &= G_3 + (A_2-I_d) (\overline{\Delta_1 X} + \overline{\Delta_2 X}) \\
 \widehat{\Xi}_3 &= (A_2-I_d) (\overline{T}_1+\overline{T}_2) (A_2-I_d)' + T_3
\end{align*}
 \item[2.] Updating: $\mathcal{L}(\Delta_2 X|Y_2,Y_1) = \mathcal{N}_d(\overline{\Delta_2 X},\overline{T}_2)$
 \begin{align*}
  \overline{\Delta_2 X} &= \widehat{\Delta_2 X} + \widehat{\Xi}_2 \tilde{B}' (\tilde{B} \widehat{\Xi}_2 \tilde{B}' + \tilde{P}_2)^{-1}(Y_2 - \tilde{B}\widehat{\Delta_2 X}) \\
  \overline{T}_2 &= \widehat{\Xi}_2 - \widehat{\Xi}_2 \tilde{B}' (\tilde{B} \widehat{\Xi}_2 \tilde{B}' + \tilde{P}_2)^{-1} \tilde{B} \widehat{\Xi}_2
 \end{align*}
\item[3.] Marginal: $\mathcal{L}(Y_3|Y_2,Y_1) = \mathcal{N}(\widehat{M}_3,\widehat{\Omega}_3)$
\begin{align*}
 \widehat{M}_3 &= \tilde{B} \widehat{\Delta_3 X} \\
 \widehat{\Omega}_3 &= \tilde{B} \widehat{\Xi}_3 \tilde{B}' + \tilde{P}_3
\end{align*}
\end{itemize}

\noindent \textbf{Proof:} We just have to prove that, conditionally on $Y_1$, $Y_2$, $\Delta_1 X$ and $\Delta_2 X$ are independent. First, we have:
\begin{equation*}
 \Delta_3 X = G_3 + A_2(\Delta_1 X + \Delta_2 X) + U_3.
\end{equation*}
Hence: 

\begin{equation*}
 \mathbb{E}(\Delta_3 X|Y_2,Y_1) = G_3 + A_2 (\mathbb{E}(\Delta_1 X|Y_1) + \mathbb{E}(\Delta_2 X|Y_2,Y_1)) = G_3 + A_2(\overline{\Delta_1 X} + \overline{\Delta_2 X}).
\end{equation*}

\noindent Let $t_1,t_2 \in \mathbb{R}^d$. Then, we can compute the characteristic function of $\Delta_1 X+\Delta_2 X$ conditionally to $Y_2$, $Y_1$: 

\begin{align*}
 \mathbb{E}\left[\exp\left(i t_1' \Delta_1 X + i t_2'\Delta_2 X\right) | Y_2,Y_1 \right] &= \mathbb{E}\left[\exp\left(i t_1' \Delta_1 X\right)|Y_2,Y_1 \right] \mathbb{E}\left[\exp\left(i t_2'\Delta_2 X| \Delta_1 X\right),Y_2,Y_1 \right] \\
 &= \exp\left(t_1' \overline{\Delta_1 X} + \frac{1}{2} t_1' \overline{T}_1\right) \times \exp\left(t_2' \overline{\Delta_2 X} + \frac{1}{2} t_2' \overline{T}_2\right).
\end{align*}

\noindent Consequently, conditionally to $Y_1$, $Y_2$, $\Delta_1 X$ and $\Delta_2 X$ are independent and $$\text{Var}(\Delta_1 X+\Delta_2 X|Y_2,Y_1) = \overline{T}_1 + \overline{T}_2.$$ \begin{flushright} $\square$ \end{flushright}

\noindent Then, the generalization to the case $k\ge 1$ is direct, leading to the Kalman filter described in Section \ref{sec:inference} for incidence data. 

\section{Practical considerations on implementation setting}
\label{sec:rem_implementation}

Let us make some remarks on practical implementation. \\

\begin{itemize}

\item Two strategies for the choice of the step-size $\alpha_m$ at a given iteration $m$ of the SAEM algorithm are combined, as recommended in (\cite{Lavielle2014}): first, denoting by $M_0$ the number of burn-in iterations, we use $\alpha_m=1$ if $m\le M_0$ to quickly converge to a neighborhood of the solution and then, $\alpha_m = \frac{1}{(m-M_0)^{\nu_0}}$ if $m>M_0$ with $\frac{1}{2} \le \nu_0\le 1$ to ensure almost sure convergence of the sequence $(\theta_m)$ to the maximum likelihood estimate of $\theta$. \\

\item An extended algorithm for non-exponential models is proposed to include fixed effects (see e.g. \cite{Debavelaere2021}). Let $\kappa$ be a fixed parameter to be estimated. First, for $m=1,\ldots,M_0$, we use the classical procedure of the SAEM algorithm, that is a mean and a variance of the parameter is estimated at each iteration as if it were a random parameter. Then, at each new iteration $m+1$, the current variance of the parameter, denoted $\omega_{\kappa}^{(m+1)}$, is updated as:  $\omega_{\kappa}^{(m+1)} = K_0 \times  \omega_{\kappa}^{(m)}$, with $0<K_0<1$. \\

\item Due to the small influence of the number of iterations in the Metropolis-Hastings procedure (see e.g. \cite{Kuhn2005}), a single iteration is used. Furthermore, if the proposal distribution is the marginal distribution $p(\mathbf{\Phi};\tilde{\theta})$, the expression of the acceptance probability is simplified as follows:

\begin{equation*}
 \rho(\mathbf{\Phi}_{m-1},\mathbf{\Phi}^{(c)}) = \text{min}\left[1, \frac{p(\mathbf{y}|\mathbf{\Phi}^{(c)};\tilde{\theta})}{p(\mathbf{y}|\mathbf{\Phi}_{m-1};\tilde{\theta})} \right].
\end{equation*}

\item A stopping criterion for the SAEM algorithm is considered. Denote by $\theta_j^{(m)}$ the $j$-th component of $\theta$ estimated at iteration $m$ of the SAEM algorithm. Then, the algorithm stops either when the criterion  $$\max_j\left(\frac{|\theta_j^{(m)}-\theta_j^{(m-1)}|}{|\theta_j^{(m)}|}\right)<\mu_0$$ is satisfied several times consecutively or when a limit of $M_{\max}$ iterations is reached. The value of $\mu_0$ is chosen sufficiently small (e.g. of the order of $10^{-3}$ or $10^{-4}$). \\

\item As the convergence of the SAEM algorithm can strongly depend on the initial guess, a simulated annealing version of SAEM (\cite{Kirkpatrick1984}) is used to escape from potential local maxima of the likelihood during the first iterations and converge to a neighborhood of the global maximum. Let $\hat{\Gamma}\left(\phi_m^{(j)}\right)$ the estimated variance of the $j$-th component of $\mathbf{\Phi_m}$ at iteration $m$ of the SAEM algorithm. Then, while $m\le M_0$, $\Gamma_m^{(j)}=\text{max}\left[\tau_0 \ \Gamma_{m-1}^{(j)},\hat{\Gamma}\left(\phi_m^{(j)}\right)\right]$ with $0<\tau_0<1$. For $m>M_0$, the usual SAEM algorithm is used to estimate the variances at each iteration (see e.g. \cite{Lavielle2014}). \\ 

\item For the initialization of the SAEM algorithm, the starting parameter values $\beta_{0}$ of the fixed effects $\beta$ are uniformly drawn from a hypercube encompassing the likely true values. The initial variances $\Gamma_0$ are chosen sufficiently large ($1$ by default).   \\

\item When the sampling intervals between observations $\Delta$ are large, the approximation of the resolvent matrix proposed in (\cite{Narci2020}), Appendix A, is used. \\

\item Concerning the KM approach, we use the Nelder-Mead method implemented in the \verb!optim! function of the \verb!R! software to maximize the approximated log-likelihood given by the Kalman filter. This requires to provide some initial values for the unknown parameters. As the optimization can be very sensitive to initialisation, $10$ different starting values are considered and the maximum value for the log-likelihood among them are chosen. The starting parameter values for the maximization algorithm are uniformly drawn from a hypercube encompassing the likely true values. \vspace{0.07cm}

\end{itemize}

\noindent For simulation studies in Section \ref{sec:datasimulations}, the tuning parameters values are chosen as: $M_0=500$, $\nu_0=0.6$, $K_0=0.87$, $\mu_0=0.001$, $M_{\max}=1000$ and $\tau_0=0.98$. Concerning the investigation of influenza outbreaks in Section \ref{sec:estimrealdata}, we chose: $M_0=5000$, $\nu_0=0.6$, $K_0=0.87$, $\mu_0=0.0001$ and $\tau_0=0.98$. The algorithm stops when the criterion is checked $100$ times successively.

\section{Estimation results for a second set of parameter values}

\subsection{Simulation settings}

We consider a second set of parameter values which induces a lower intrinsic variability between epidemics. As for the first set of values, we consider two settings (denoted respectively (i) and (ii)) corresponding to two levels of inter-epidemic variability (resp. high and moderate): 
\begin{itemize}
  \item Setting (i): $\beta=(0.58,1.10,1.45,-2.20)'$ and $\Gamma=\text{diag}(0.47^2,1.5^2,0.75^2)$ corresponding to $\mathbb{E}\left(R_{0,1:U}\right)=3$, $CV_{R_0} = 33 \%$; $d=3$; $\mathbb{E}\left(p_{1:U}\right)\approx 0.74$, $CV_p \approx 31 \%$; $\mathbb{E}\left(i_{0,1:U}\right)\approx0.12$, $CV_{i_0} \approx 66 \%$. \\
   \item Setting (ii): $\beta=(0.66,1.10,1.45,-2.2)'$ and $\Gamma=\text{diag}(0.25^2,0.9^2,0.5^2)$ corresponding to $\mathbb{E}\left(R_{0,1:U}\right)=3$, $CV_{R_0} = 17 \%$; $d=3$; $\mathbb{E}\left(p_{1:U}\right)\approx0.78$, $CV_p \approx 18 \%$; $\mathbb{E}\left(i_{0,1:U}\right)\approx0.11$, $CV_{i_0} \approx 45 \%$. 
\end{itemize}

\subsection{Point estimates and standard deviation for inferred parameters}

\noindent Tables \ref{exp2_CV2} and \ref{exp2_CV1} show the estimates of the expectation and standard deviation of the random effects $\phi_{u}$, computed from the estimations of $\beta$ and $\Gamma$ using functions $h$ defined in (\ref{phiSIR}), for settings (i) and (ii). For each parameter, the reported values are the mean of the $J=100$ parameter estimates $\phi_{u,j}$, $j\in \{1,\ldots,J\}$, and their standard deviations in brackets. 

{\setlength{\tabcolsep}{6pt}
\begin{table}[H]
\begin{center}
\caption{Estimates for setting (i): high inter-epidemic variability. For each combination of $(\overline{n},U)$ and for each model parameter (defined in the first line of the table), point estimates and precision are calculated as the mean of the $J=100$ individual estimates and their standard deviations (in brackets).} \label{exp2_CV2}
\small
\begin{tabular}{cc|ccccccc}
Parameters & &  $\mathbb{E}\left(R_{0,u}\right)$ & $d$  & $\mathbb{E}\left(p_{u}\right)$  & $\mathbb{E}\left(i_{0,u}\right)$ & 
$\text{sd}\left(R_{0,u}\right)$ & $\text{sd}\left(p_{u}\right)$ & $\text{sd}\left(i_{0,u}\right)$ \\
\hline
True values & & \textbf{3.000} & \textbf{3.000} & \textbf{0.739} & \textbf{0.119} & \textbf{1.000} & \textbf{0.226} & \textbf{0.079} \\
 \hline
  & & & & & & & &  \\
 $\overline{n}=20$ & $U=20$  & 3.085  &  2.889  &  0.758  &  0.111  &  1.477  &  0.205  &  0.075  \\
  & &  (\textit{0.460}) & (\textit{0.205}) & (\textit{0.060}) & (\textit{0.016}) & (\textit{0.666}) & (\textit{0.036}) & (\textit{0.018}) \\
    & & & & & & & &  \\
 & $U=50$  &  3.152  &  2.926  &  0.761  &  0.111  &  1.509  &  0.199  &  0.075  \\
 & &  (\textit{0.360}) & (\textit{0.170}) & (\textit{0.049}) & (\textit{0.011}) & (\textit{0.457}) & (\textit{0.025}) & (\textit{0.012}) \\
   & & & & & & & &  \\
  & $U=100$   & 3.116  &  2.904  &  0.765  &  0.111  &  1.517  &  0.200  &  0.077  \\
 &   &  (\textit{0.307}) & (\textit{0.152}) & (\textit{0.046}) & (\textit{0.008}) & (\textit{0.366}) & (\textit{0.018}) & (\textit{0.009}) \\
   & & & & & & & &  \\
  \hline
  & & & & & & & &  \\
$\overline{n}=100$ & $U=20$   & 2.929  &  2.932  &  0.742  &  0.116  &  1.124  &  0.212  &  0.075  \\
  & & (\textit{0.263}) & (\textit{0.144}) & (\textit{0.047}) & (\textit{0.016}) & (\textit{0.332}) & (\textit{0.029}) & (\textit{0.017}) \\
    & & & & & & & &  \\
 & $U=50$  &  3.002  &  2.973  &  0.749  &  0.116  &  1.186  &  0.207  &  0.075  \\
 & &  (\textit{0.242}) & (\textit{0.116}) & (\textit{0.031}) & (\textit{0.012}) & (\textit{0.315}) & (\textit{0.022}) & (\textit{0.011}) \\
   & & & & & & & &  \\
  & $U=100$   & 2.952  &  2.942  &  0.751  &  0.115  &  1.159  &  0.212  &  0.075  \\
 &   &  (\textit{0.148}) & (\textit{0.090}) & (\textit{0.022}) & (\textit{0.008}) & (\textit{0.155}) & (\textit{0.018}) & (\textit{0.007}) \\
\end{tabular}
\end{center}
\end{table}
}

\vspace{-1.1cm}

{\setlength{\tabcolsep}{6pt}
\begin{table}[H]
\begin{center}
\caption{Estimates for setting (ii): moderate inter-epidemic variability. For each combination of $(\overline{n},U)$ and for each model parameter (defined in the first line of the table), point estimates and precision are calculated as the mean of the $J=100$ individual estimates and their standard deviations (in brackets).} \label{exp2_CV1}
\small
\begin{tabular}{cc|ccccccc}
Parameters & &  $\mathbb{E}\left(R_{0,u}\right)$ & $d$  & $\mathbb{E}\left(p_{u}\right)$  & $\mathbb{E}\left(i_{0,u}\right)$ & 
$\text{sd}\left(R_{0,u}\right)$ & $\text{sd}\left(p_{u}\right)$ & $\text{sd}\left(i_{0,u}\right)$ \\
\hline
True values & & \textbf{3.000} & \textbf{3.000} & \textbf{0.777} & \textbf{0.109} & \textbf{0.500} & \textbf{0.143} & \textbf{0.049} \\
 \hline
  & & & & & & & &  \\
 $\overline{n}=20$ & $U=20$  & 3.183  &  3.051  &  0.771  &  0.106  &  0.811  &  0.128  &  0.046  \\
  & &  (\textit{0.292}) & (\textit{0.164}) & (\textit{0.046}) & (\textit{0.012}) & (\textit{0.321}) & (\textit{0.029}) & (\textit{0.011}) \\
    & & & & & & & &  \\
 & $U=50$  &  3.201  &  3.050  &  0.765  &  0.106  &  0.874  &  0.132  &  0.048  \\
 & &  (\textit{0.208}) & (\textit{0.116}) & (\textit{0.035}) & (\textit{0.008}) & (\textit{0.241}) & (\textit{0.018}) & (\textit{0.007}) \\
   & & & & & & & &  \\
  & $U=100$   & 3.232  &  3.068  &  0.765  &  0.106  &  0.906  &  0.132  &  0.048  \\
 &   &  (\textit{0.189}) & (\textit{0.103}) & (\textit{0.028}) & (\textit{0.005}) & (\textit{0.212}) & (\textit{0.013}) & (\textit{0.005}) \\
   & & & & & & & &  \\
 \hline
  & & & & & & & &  \\
$\overline{n}=100$ & $U=20$   & 3.037  &  3.051  &  0.770  &  0.110  &  0.563  &  0.135  &  0.046  \\
  & &  (\textit{0.169}) & (\textit{0.100}) & (\textit{0.037}) & (\textit{0.012}) & (\textit{0.206}) & (\textit{0.026}) & (\textit{0.011}) \\
    & & & & & & & &  \\
 & $U=50$  & 3.064  &  3.055  &  0.764  &  0.110  &  0.632  &  0.139  &  0.048  \\
 & &  (\textit{0.117}) & (\textit{0.080}) & (\textit{0.023}) & (\textit{0.009}) & (\textit{0.142}) & (\textit{0.016}) & (\textit{0.007}) \\
   & & & & & & & &  \\
  & $U=100$   & 3.059  &  3.057  &  0.768  &  0.110  &  0.619  &  0.141  &  0.048  \\
 &   &  (\textit{0.088}) & (\textit{0.061}) & (\textit{0.019}) & (\textit{0.005}) & (\textit{0.094}) & (\textit{0.013}) & (\textit{0.004}) \\
\end{tabular}
\end{center}
\end{table}
}

\noindent As for the first set of parameters values, all point estimates are closed to the true values. The standard error of the estimates decreases when the number of epidemics $U$ and the number of observations $\bar{n}$ increases, whereas the bias is only sensitive to $\bar{n}$ (bias decreasing when $\bar{n}$ increasing). \\

\begin{figure}[H]
	\includegraphics[width=0.95\textwidth,height=10.5cm]{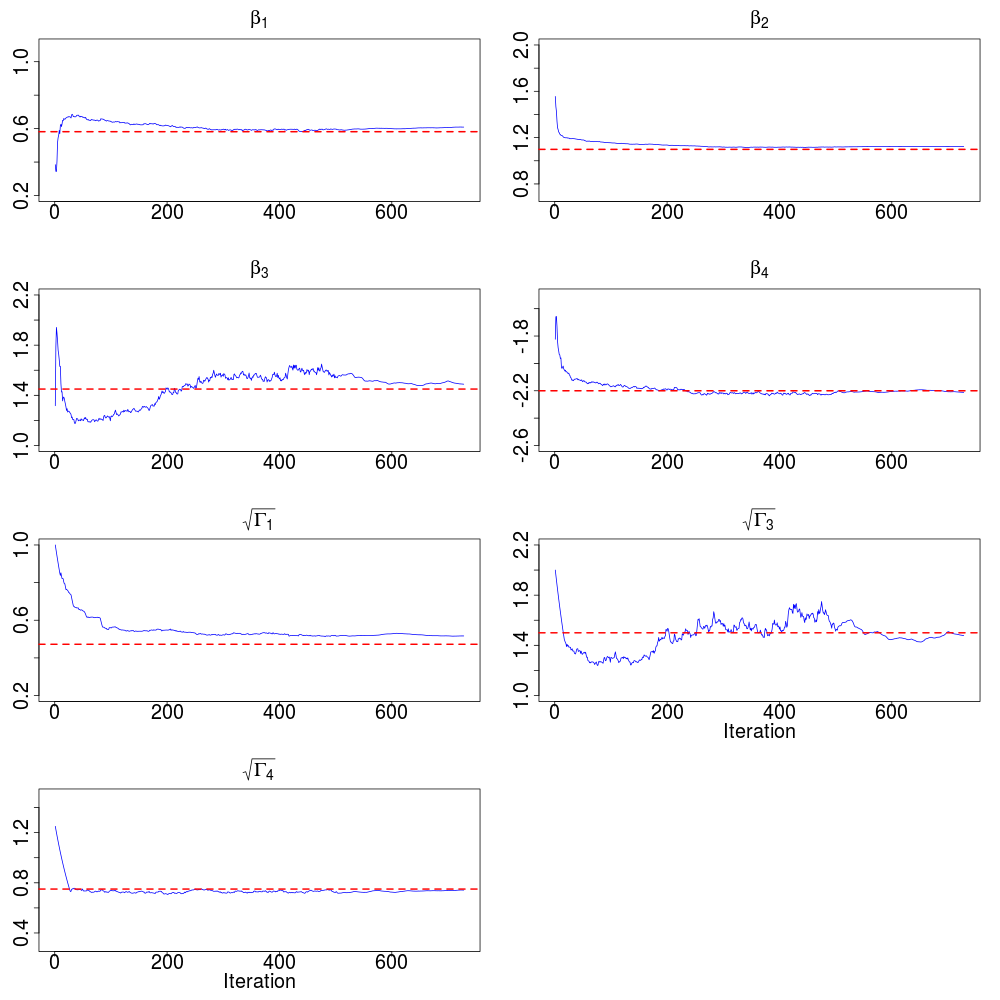}\\
	\caption{Convergence graphs of the SAEM algorithm for estimates of $\beta=(\beta_1,\beta_2,\beta_3,\beta_4)$ and $\text{diag}(\Gamma)=(\Gamma_1,\Gamma_3,\Gamma_4)$. Setting (i) with $U=100$ and $\bar{n}=100$. Parameter values at each iteration of the SAEM algorithm (plain blue line) and true values of model parameters (dotted red line).}\label{CV_exp2}
\end{figure}

\noindent For a given data set, Figure \ref{CV_exp2} displays convergence graphs for model parameters in setting (i) with $U=100$ and $\bar{n}=100$.

\noindent We notice that all model parameters converge towards their true value.

\end{document}